\documentclass[aps,prl,showpacs,twocolumn,superscriptaddress,longbibliography,amsmath,amssymb]{revtex4-1}
\usepackage[pdftex]{graphicx}
\usepackage{color}
\usepackage[pdftex,bookmarks,colorlinks,breaklinks]{hyperref}
\hypersetup{linkcolor=red,citecolor=blue,urlcolor=blue}
\hypersetup{pdftoolbar=true,pdfpagemode=UseNone,pdfstartview=FitH, colorlinks=true,extension=}
\usepackage{braket}
\usepackage{dsfont}

\usepackage{soul}

\newcommand{\be}{\begin{equation}}
\newcommand{\ee}{\end{equation}}

\begin{document}

\title{Optimal Targeted Mode Transport in Complex Wave Environments: A Universal Statistical Framework}

\author{Cheng-Zhen Wang}
\altaffiliation{These authors contributed equally to this work}
\affiliation{Wave Transport in Complex Systems Lab, Department of Physics, Wesleyan University, Middletown, CT-06459, USA}
\author{John Guillamon}
\altaffiliation{These authors contributed equally to this work}
\affiliation{Wave Transport in Complex Systems Lab, Department of Physics, Wesleyan University, Middletown, CT-06459, USA}
\author{Ulrich Kuhl}
\affiliation{Wave Transport in Complex Systems Lab, Department of Physics, Wesleyan University, Middletown, CT-06459, USA}
\affiliation{Université Côte d’Azur, CNRS, Institut de Physique de Nice (INPHYNI), 06200, Nice, France}
\author{Matthieu Davy}
\affiliation{Université de Rennes, CNRS, IETR- UMR 6164; F-35000 Rennes, France}
\author{Mattis Reisner}
\affiliation{Wave Transport in Complex Systems Lab, Department of Physics, Wesleyan University, Middletown, CT-06459, USA}
\author{Arthur Goetschy}
\email[]{ arthur.goetschy@espci.psl.eu}
\affiliation{ESPCI Paris, PSL University, CNRS, Institut Langevin, Paris, France}
\author{Tsampikos Kottos}
\email[]{tkottos@wesleyan.edu}
\affiliation{Wave Transport in Complex Systems Lab, Department of Physics, Wesleyan University, Middletown, CT-06459, USA}

\begin{abstract}
Recent advances in the field of structured waves have resulted in sophisticated coherent wavefront shaping schemes that provide unprecedented control of waves in various complex settings. These techniques exploit multiple scattering events and the resulting interference of wave paths within these complex environments. Here, we introduce the concept of targeted mode transport (TMT), which enables energy transfer from specific input channels to designated output channels in multimode wave-chaotic cavities by effectively engaging numerous cavity modes. We develop a statistical theory that provides upper bounds on optimal TMT, incorporating operational realities such as losses, coupling strengths and the accessibility of specific interrogating channels. The theoretical predictions for the probability distribution of TMT eigenvalues are validated through experiments with microwave chaotic networks of coaxial cables as well as two-dimensional and three-dimensional complex cavities. These findings have broad implications for applications ranging from indoor wireless communications to imaging and beyond.
\end{abstract}

\maketitle

\section{Introduction}
The inherent complexity and acute sensitivity generated by multiple scattering and the resulting interference of numerous ray paths in electromagnetic reverberant environments are often viewed as significant challenges for resilient indoor wireless communications that aim to transmit information or 
energy to specific targeted channels or areas~\cite{carneiro2022study,de2021electromagnetic}. To overcome 
this challenge, various coherent wavefront shaping (CWS) schemes have been proposed, and successfully 
implemented, in diverse complex settings~\cite{cao2022shaping,gigan2022roadmap, Fei2024,chen2020perfect,sweeney2020theory,jiang2024coherent,wang2024nonlinearity,goicoechea2024detecting,guo2023singular,
sol2023reflectionless}. These protocols, however, are inherently deterministic, requiring detailed knowledge of the scattering processes associated with the specific structure of the complex medium under study. This dependence limits their practicality for wireless communication systems, where small temporal variations in the enclosure configuration, coupling to interrogating or targeted antennas, or operating frequency, can drastically alter the scattering process.

In practical scenarios, the pursuit of wavefronts achieving 100\% targeted mode transport (TMT) efficiency to a specified set of outgoing channels—distinct from the injected ones—is not only exceptionally challenging but often irrelevant. Instead, a more realistic approach would recognize the sensitivity of the scattering environment to small perturbations and adopt a statistical framework to describe the probabilistic conditions for maximal targeted transmission efficiency~\cite{chong2011hidden,goetschy2013filtering,popoff2014coherent,hsu2017correlation,li2017random,li2018statistical,boucher2021full,bender22, mcintosh2024delivering}. Such an approach would establish optimal bounds for TMT while accounting for operational realities, such as cavity losses~\cite{kuhl2005direct,gopar2007chaotic,baez2008absorption,liew2014transmission}, coupling efficiencies between the scattering domain and interrogating or targeted antennas~\cite{brouwer1995generalized,savin2001reducing,Fyodorov_2017}, and incomplete channel control~\cite{goetschy2013filtering,mcintosh2024delivering}. The latter often arises from experimental constraints on generating input channel states, such as limited apertures for the injected waves.

\begin{figure*}
\centering
\includegraphics[width=\linewidth]{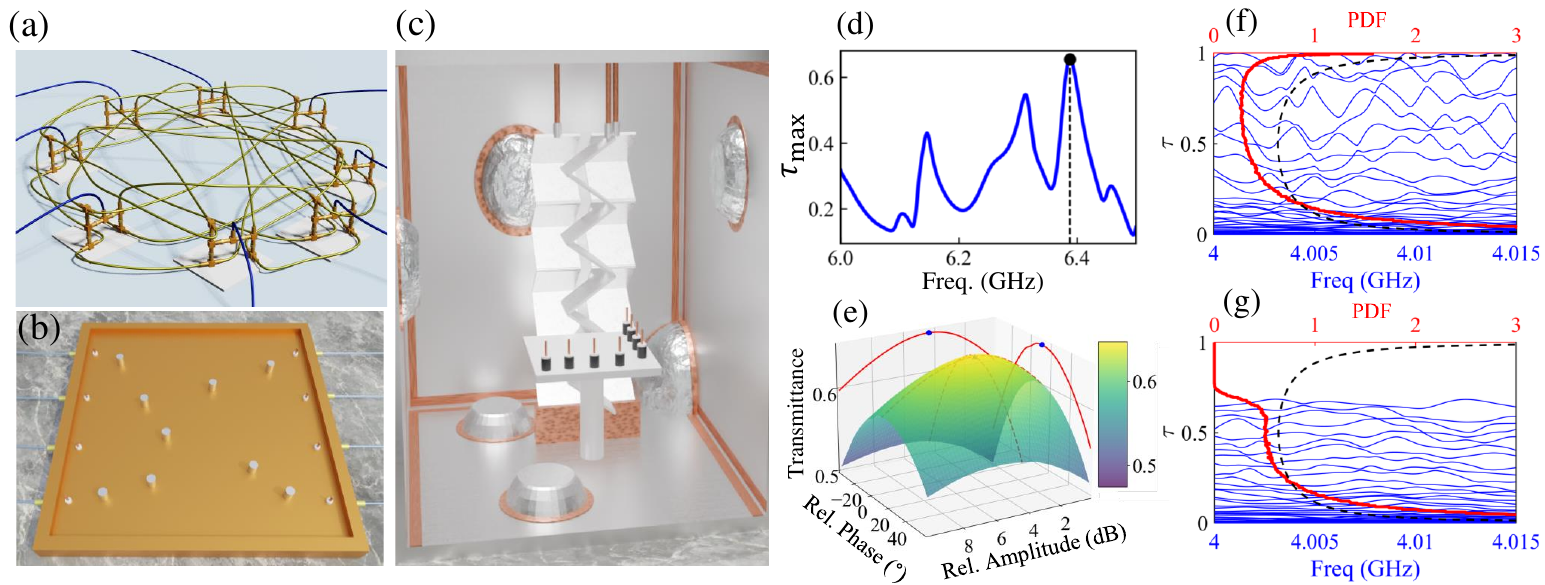}
\caption{ \label{Fig1}
(a) Quasi-one-dimensional complex network of coaxial cables. (b) Two-dimensional microwave chaotic cavity. (c) Three-dimensional reverberation chamber. (d) Experimental maximum eigenvalue of the TMT matrix as a function of input wave frequency for a network with $M=4$  channels. The TMT process involves two input channels and one targeted output channel. The frequency at which the overall maximum TMT occurs within the analyzed frequency range 
is $f \approx 6.39$ GHz, indicated by the dashed vertical line. (e) Corresponding transmittance of the network used in subfigure (d) as a function of the relative amplitude and phase of a two-port injected wavefront at the TMT frequency $f\approx 6.39$ GHz. (f) Simulated TMT eigenvalues versus frequency (blue lines) for a network of $300$ vertices coupled to $M=80$ channels with $M_{\rm in}=40$ input channels and $M_{\rm tar}=40$ targeted channels. The coupling parameter is $\Gamma \approx 0.49$.  Each vertex is coupled to $6$ other, randomly chosen vertices, with bond lengths randomly sampled from a uniform distribution in the range [$0.1$ m,  $0.4$ m]. The probability density distribution (red line) is generated over frequencies in the range [$3.5$ GHz – $4.5$ GHz]. The bimodal distribution is also shown for comparison (black dashed line). (g) Same plot as (e), but with uniform loss added to the cables, represented by an imaginary part of the refractive index $n_i = 2\times 10^{-4}$.
}
\end{figure*}

Here, we challenge the conventional notion that complex environments inherently hinder the development of efficient energy-targeting schemes. Instead, we exploit the intricate electromagnetic interferences generated within reverberant multimode cavities to derive a \emph{universal} statistical theory. This theory predicts the distribution of TMT eigenvalues for wavefronts emitted from multi-element sources to specified target channels. Our diagrammatic approach incorporates realistic considerations such as internal losses, imperfect coupling between antennas and the scattering domain, and incomplete access to the physical scattering matrix. 

The theoretical predictions show excellent agreement with \textit{ab initio} wave propagation simulations and offer valuable insights into experimental results obtained from various multimode microwave platforms. These include quasi-one-dimensional complex networks of coaxial cables, two-dimensional chaotic cavities, and three-dimensional reverberation chambers, as illustrated in Fig.~\ref{Fig1}(a,b,c). Our framework identifies the macroscopic parameters that govern TMT efficiency, regardless of the microscopic details of the medium. Moreover, it provides explicit bounds for optimal wave control and uncovers counterintuitive conditions---such as the interplay between the number of controlled modes and their coupling strength---that maximize efficiency.

The universal nature of our results extends their applicability beyond indoor wireless communications to a wide range of physical platforms dominated by chaotic ray dynamics. Examples include optical systems, where spatial light modulators enable the transmission of information through opaque media, as well as satellite communications utilizing microwaves or mid-infrared waves propagating through the atmosphere. Additional applications span ultrasonic imaging in multi-scattering media, seismic wave analysis, and more.

\section{Physical Platforms for TMT}

The experimental platforms used for the statistical analysis of TMTs included complex networks, two-dimensional (2D), and three-dimensional (3D) chaotic cavities, as shown in Figs.~\ref{Fig1}(a,b,c), respectively. Complex networks of coupled coaxial microwave cables (Fig.~\ref{Fig1}(a)) have proven to be both simple and versatile platforms for experimentally demonstrating and theoretically analyzing wave phenomena in systems with underlying classical chaotic dynamics~\cite{kottos2001,kottos2003,pluhavr2013,pluhavr2014}. These networks are frequently used as models for mesoscopic quantum transport, sound propagation, and electromagnetic wave behavior in complex interconnected structures such as buildings, ships, and aircrafts~\cite{hurt2000mathematical,kuchment2002graph,kuchment2004quantum,berkolaiko2013introduction}. The scattering matrix $S$ was measured by connecting the network to transmission lines, which were coupled to the ports of a vector network analyzer (VNA). To enable statistical processing of TMTs, multiple network configurations were generated by scanning the interrogation frequency over the range [$1.5$ GHz, $4.5$ GHz] and systematically exploring all possible configurations among the available channels for each TMT scenario.

To further test our theory on the statistical properties of TMTs in complex systems, we conducted additional experiments using 2D chaotic cavities (Fig.~\ref{Fig1}(b)). Metallic cylinders were placed at random positions inside the cavity, and the scattering matrix was measured using a VNA connected to matched coax-to-waveguide antennas attached to the cavity. A statistical ensemble has been produced by tuning the frequency in the range [$8$ GHz, $15$ GHz] and creating all possible configurations among the available channels. Finally, we analyzed the TMT statistics in 3D chaotic enclosures (reverberation chambers, Fig.~\ref{Fig1}(c)). Here, the scattering matrix was measured using commercial WiFi antennas. The reverberation chamber was equipped with two mechanical stirrers, one horizontal and one vertical, allowing us to generate a random ensemble of scattering matrix configurations. Additional statistics were generated by measuring the $S$-matrix over the frequency range [$2$ GHz, $3$ GHz].

In all these systems, the TMT process is characterized by the efficient coupling of a specific subset of scattering channels, controlled by another subset among the $M$ available channels. The portion of the total scattering matrix describing the TMT process is given by $\tilde{S} = P_{\rm tar} S P_{\rm in}$,
where $P_{\rm in}$ and $P_{\rm tar}$ are $M\times M_{\rm in}$ and $M_{\rm tar} \times M$ projection matrices. These matrices define the subspaces of $M_{\rm in}$ controlled input channels and $M_{\rm tar}$ targeted output channels, respectively. In this paper, we require that {\it these subspaces are distinct}, as is typical in wireless communication protocols, which is expressed by the orthogonality condition $P_{\rm tar} \cdot P_{\rm in} = 0$.

For an incident wavefront $|\psi_{\rm in} \rangle$ confined to the $P_{\rm in}$-subspace, the outgoing signal measured in the $P_{\rm tar}$-subspace after propagation through the complex multimode cavity is $\langle \psi_{\rm in} | \tilde{S}^\dagger \tilde{S} | \psi_{\rm in} \rangle$.
Consequently, the eigenvalues $\tau$ of the TMT matrix $T = \tilde{S}^\dagger \tilde{S}$ govern the efficiency of the process. Specifically, the extremal eigenvalues (and their corresponding eigenvectors) represent the maximum and minimum achievable TMT processes in such setups, enabling the design of wavefront schemes with extreme transport characteristics.

An example of a TMT wavefront for the system in Fig.~\ref{Fig1}(a), connected to $M=4$ antennas, is illustrated in Figs.~\ref{Fig1}(d-e).  
Here, the wavefront is designed to inject an incident wave into the network via antennas $\alpha=1,2$, aiming to maximize the transmittance to the targeted port $\alpha=3$. Due to Ohmic losses in the coaxial cables and non-ideal coupling between the antennas and the network, the maximum achievable transmittance for this TMT process is $\tau_{\rm max} \approx 0.65$, occurring at $f \approx 6.39\,\mathrm{GHz}$ (see Fig.~\ref{Fig1}(d)). This is achieved by injecting a wavefront at ports $\beta=1,2$ with the relative amplitude and phase determined by the eigenvector components of the $2 \times 2$ TMT matrix $T$, as shown in Fig.~\ref{Fig1}(e).  

The number of controlled ($M_{\rm in}$) and targeted ($M_{\rm tar}$) channels, along with imperfect coupling and inherent losses, impose an upper limit on the efficiency of optimal TMT processes. This challenge is particularly pronounced in systems with many channels ($M \gg 1$), where the statistical behavior of the eigenvalues of the TMT matrix is governed by complex correlations, as illustrated by the blue lines in Figs.~\ref{Fig1}(f,g). To gain deeper insight into this phenomenon, we conducted extensive simulations of wave dynamics in random networks and cavities and developed an analytical framework to describe the eigenvalue distribution of the TMT matrix, as detailed below.

\begin{figure*}
\centering
\includegraphics[width=\linewidth]{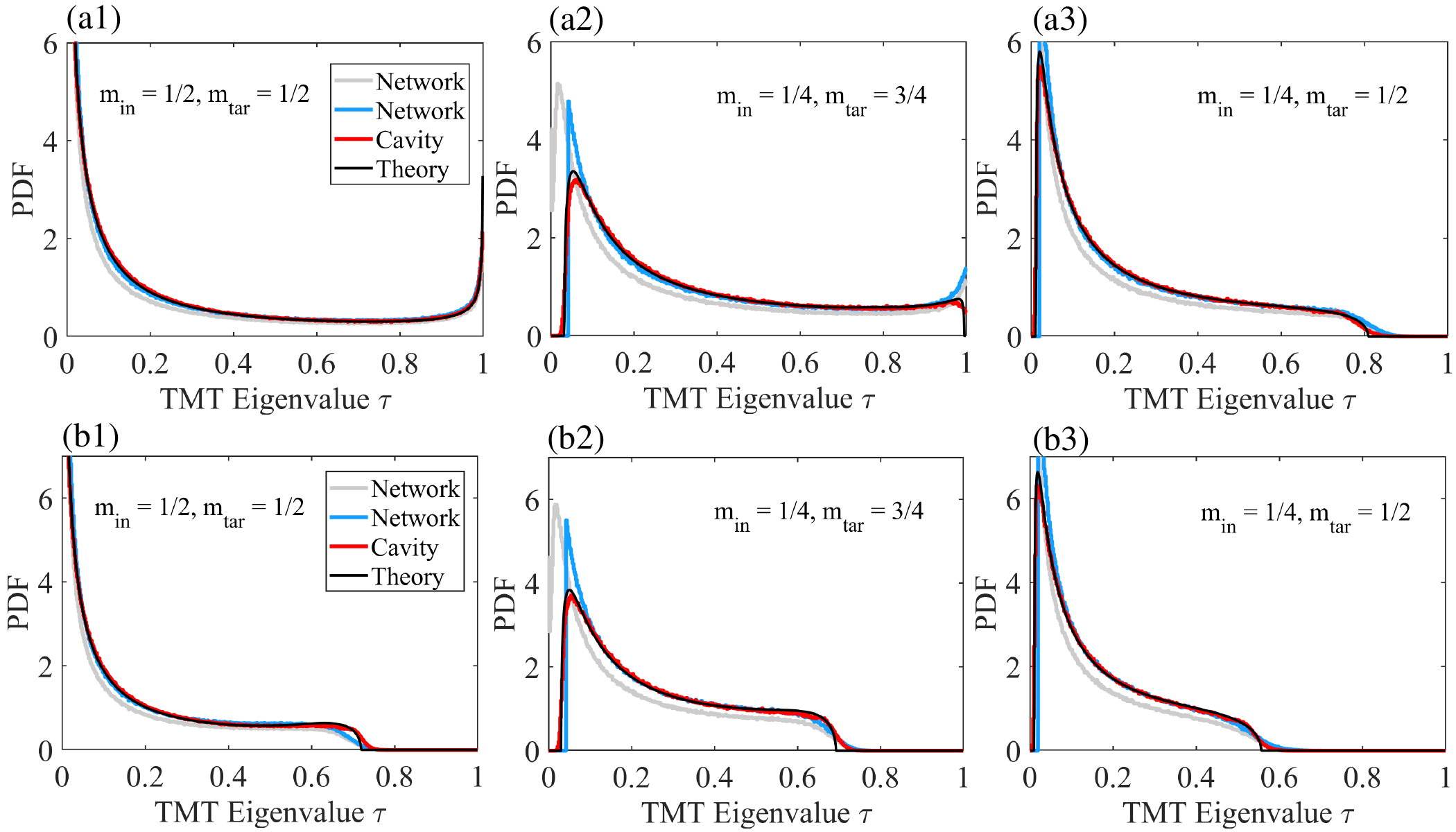}
\caption{ \label{Fig2}
Probability density function (PDF) of TMT eigenvalues $\tau$ evaluated from wave simulations of a complex network  (grey lines), and a cavity model (red lines). The blue lines indicate the PDF of the TMT eigenvalues of the network system after removing the lowest 18\% of eigenvalues, which are associated with suppressed transmission due to semiclassical effects (scarring) and other localization phenomena specific to the network model~\cite{schanz2003,wang2024bound}. 
The ensemble has been generated from random configurations of cable lengths uniformly distributed in the interval [$0.1$ m, $0.4$ m] over the frequency range [$3.5$ GHz, $4.5$ GHz] for the networks, and from random Gaussian matrices for the cavity model. Panels (a1)–(a3) represent lossless systems with varying input ($m_{\rm in}$) and output ($m_{\rm tar}$) channel ratios. Panels (b1)–(b3) show the PDF in the presence of losses for the same systems. For the complex network, the imaginary part of the cable refractive index is $2\times 10^{-4}$, corresponding to absorption $A\approx 0.13$. For the lossy cavity model, the loss in the diagonal elements of the Hamiltonian matrix is $\gamma'\approx 0.016$ corresponding to the same $A\approx 0.13$. In all cases, the complex network consists of $300$ vertices, and is attached to $80$ TLs. Each vertex is randomly coupled to six others. The cavity model consists of $300$ modes and is attached to $80$ TLs. Both systems are characterized by a coupling parameter $\Gamma \approx 0.49$.  In all subfigures, the predictions of the diagrammatic theory are shown (black lines). In (b1)-(b3), the absorption factor is $a\approx0.17$.
}
\end{figure*}

\section{Statistical theory and wave simulations}

To efficiently solve the wave equation for random networks and multimode cavities, we consider systems with $N \gg 1$ modes (or vertices) coupled to $M$ transmission lines (TLs) that are used to inject and receive monochromatic waves of frequency $\omega$. The coupling is characterized by a set of parameters $\gamma_\alpha$ ($\alpha = 1, \dots, M$). The incident $|\psi_{\rm in}\rangle$ and outgoing $|\psi_{\rm out}\rangle$ waves are related by the equation $|\psi_{\rm out}\rangle = S |\psi_{\rm in}\rangle$. For both complex networks and chaotic cavities, the scattering matrix $S$ can be expressed as (see Methods) 
\begin{align}
\quad S(\omega) = -\hat{I} + 2 i D^\dagger \frac{1}{{\cal M}(\omega)+iDD^{\dagger}}D,
\label{Eq1}
\end{align}
where $D$ is the coupling matrix with elements $D_{n\alpha} = \sqrt{\gamma_\alpha}$ for a mode $n$ coupled to a TL, and $D_{n\alpha}=0$ otherwise. The coupling strength with the TLs is also characterized by the parameters $\Gamma_\alpha = 1 - |\langle S_{\alpha \alpha}\rangle|^2$, with $\Gamma_\alpha=1$ indicating a perfect (impedance-matched) coupling. Finally, ${\cal M}(\omega)$ represents the internal Hamiltonian dynamics within the complex isolated ($\gamma_\alpha=0$) system.

Complex networks and cavities differ in their internal matrix ${\cal M}(\omega)$ and coupling matrix $D$. For complex networks, ${\cal M(\omega)}$ explicitly depends on the adjacency matrix of the network (see Methods), which need not be fully connected, as well as on the losses introduced through the imaginary part of the refractive index within the cables. The matrix $D$ is defined such that $\gamma_{\alpha} = 1$ when TL $\alpha$ is attached to a node.

In contrast, for cavities, ${\cal M}(\omega) = \omega - H_0$, where the $N \times N$ effective Hamiltonian $H_0$ represents the wave propagation inside the isolated cavity. In general, $H_0$ is non-Hermitian due to (Ohmic or radiative) losses within the cavity. These losses are modeled by introducing an imaginary part $\gamma'$ (loss rate) in the diagonal elements of $H_0$. For chaotic cavities, $H_0$ is statistically modeled as a random matrix whose elements are  drawn from a Gaussian distribution, with the variance given by $\langle (H_0)_{nm}^2 \rangle = \frac{1}{N}(1 + \delta_{nm})$ (in units of a central frequency $\omega_0$ around which the measurements are performed). On the other hand, the coupling considered in this work is of the form $\gamma_{\alpha} = \gamma$ for all TLs, so that $\Gamma = 1 - \frac{(1-\gamma)^2}{(1+\gamma)^2}$ \cite{FS97}.

In Figs.~\ref{Fig1}(f,g), we present numerical results for the distribution ${\cal P}(\tau)$ of 
TMT eigenvalues (red line) for a network operating in the range [$3.5$ GHz, $4.5$ GHz]. The network 
consists of $N=300$ vertices, each randomly connected to $6$ others on average via coaxial cables with random lengths 
uniformly chosen in the range $[0.1~\mathrm{m}, 0.4~\mathrm{m}]$. The total numbers of controlled and targeted channels are $M_{\rm in}=40$ and $M_{\rm tar}=40$, selected from $M=80$ available channels. In this example, the imperfect coupling to the TLs is characterized by $\Gamma \approx 0.49$. 
The distribution ${\cal P}(\tau)$ is compared with the well-known bimodal prediction ${\cal P}(\tau) = \frac{1}{\pi} \frac{1}{\sqrt{\tau (1 - \tau)}}$ (black dashed line), which corresponds to a symmetric TMT process ($M_{\rm in} = M_{\rm tar} \gg 1$) under conditions of perfect coupling ($\Gamma = 1$) and no absorption~\cite{beenakker97}.
Fig.~\ref{Fig1}(f) demonstrates that imperfect coupling skews the distribution towards smaller $\tau$ values, while preserving the maximum transmittance $\tau_{\rm max} \approx 1$. Conversely, the impact of absorption, shown in Fig.~\ref{Fig1}(g), compresses the spectrum of eigenvalues, pushing the entire distribution toward smaller $\tau$ values. This leads to an unimodal structure in ${\cal P}(\tau)$ and a reduced maximum transmittance, $\tau_{\rm max} < 1$.

To gain deeper insight into the parameters governing the distribution ${\cal P}(\tau)$ in the limit where $M_{\rm in}, M_{\rm tar} \gg 1$, we turn to an analytical approach. In the simplest case of perfect coupling ($\Gamma = 1$) and no absorption, all scattering channels of the matrix $S$ are statistically equivalent. Under these conditions, the filtered random matrix (FRM) theory~\cite{goetschy2013filtering}, which has been successfully implemented in various disordered systems in recent years~\cite{popoff2014coherent,hsu2017correlation, boucher2021full, bender22, bender2022_2, mcintosh2024delivering}, can be applied directly to the matrix $S$ (see Supplementary Section II.B). In the regime of a small number of controlled channels ($m_{\rm in} \ll 1$), the distribution ${\cal P}(\tau)$, parametrized by the ratios $m_{\rm in}=M_{\rm in}/M$ and $m_{\rm tar}=M_{\rm tar}/M$, is concentrated around its mean $\langle \tau \rangle = m_{\rm tar}$ with a finite support $[\tau^-, \tau^+]$, where $\tau^- > 0$ and $\tau^+ < 1$. As $m_{\rm in}$ increases, the distribution begins to spread and eventually reaches the upper limit $\tau^+ = 1$ when the complementary channel constraint (CCC) $m_{\rm in} + m_{\rm tar} = 1$ is satisfied. Under this condition, $\tau^- > 0$ for all $m_{\rm tar}$ except for the symmetric case $m_{\rm in} = m_{\rm tar} = 1/2$, where the bimodal distribution is recovered.

The more complex and realistic scenario of imperfect coupling and finite absorption cannot be addressed using the same FRM formalism. This is primarily because selecting a subset of injection channels disrupts the equivalence among the outgoing channels of the scattering matrix $S$ when $\Gamma \neq 1$. In particular, the matrices $P_{\rm in}$ and $P_{\rm tar}$ are no longer statistically equivalent under the orthogonality constraint $P_{\rm tar} \cdot P_{\rm in} = 0$. 
To address these scenarios, which frequently arise in wireless communication frameworks, a diagrammatic approach is required~\cite{brouwer96}. 
In the limit $M_{\rm in}, M_{\rm tar} \gg 1$, we derive the distribution as ${\cal P}(\tau) = -\frac{1}{\pi} \lim_{\eta \to 0^+} \text{Im}[g(\tau + i\eta)]$,
where the resolvent $g(z)$ of the TMT operator $T$ is expressed as
\begin{align}
\label{Eq2}
g(z) = \frac{1}{z} \frac{1 - (1 - \Gamma)\Sigma_{\rm in}\Sigma_{\rm tar}}{1 - (1 - \Gamma)\Sigma_{\rm in}\Sigma_{\rm tar} - \Gamma\Sigma_{\rm in}/\sqrt{z}}.
\end{align}
The terms $\Sigma_{\rm in}$ and $\Sigma_{\rm tar}$ are complex self-energy elements that encapsulate the effects of multiple scattering within the complex medium. These terms account for partial reflections induced by the effective barriers at the interfaces between the inner medium and the transmission lines, as well as absorption effects during propagation. They are determined as solutions of two coupled nonlinear equations (see Methods and Supplementary Sections II.A and II.C), only parameterized by the channel ratios $m_{\rm in}$ and $m_{\rm tar}$, the coupling strength $\Gamma$ and the absorption factor $a = 4(N/M)\gamma'$ . The latter is related to the total absorption $A\equiv 1-\langle \text{Tr}(S^{\dagger}S)\rangle/M$ as $A= a/(1+a/\Gamma)$ (see Supplementary Section II.D).

The analytical predictions for ${\cal P}(\tau)$ based on Eq.~(\ref{Eq2}) are shown as black lines in Fig.~\ref{Fig2} for various $m_{\rm in}, m_{\rm tar}$ configurations. For comparison, the same figure also includes simulation results for a complex microwave network (blue lines) and a 2D chaotic cavity (red lines), both coupled to $M = 80$ TLs. Subfigures Figs.~\ref{Fig2}(a1-a3) correspond to lossless systems, while Figs.~\ref{Fig2}(b1-b3) account for uniform losses. The microwave network simulations were performed with the same parameters as in Figs.~\ref{Fig1}(f,g), except for $M_{\rm in}$ and $M_{\rm tar}$.
To demonstrate the universality of our predictions, which depend only on a few macroscopic parameters and not on the microscopic details of the system, we also analyzed chaotic cavities with the same coupling parameter $\Gamma=1-|\langle S_{\alpha,\alpha}\rangle|^2\approx 0.49$. Similarly, for the lossy case, the loss rate $\gamma' = 0.016$ in the chaotic cavity was tuned to match the total absorption, $A\approx 0.13$, of the network. In the network simulations, Ohmic losses were modeled by including an imaginary part in the refractive index, $n_i = 2\times 10^{-4}$.

Figure~\ref{Fig2} demonstrates that random networks and chaotic cavities exhibit the same distribution ${\cal P}(\tau)$ for the non-zero eigenvalues, despite their structural differences. This agreement is particularly striking when small TMT eigenvalues, associated with localization effects in sparsely connected networks (i.e., networks with low vertex valency) and system-specific phenomena such as scars, which can inhibit the development of fully ergodic dynamics \cite{pluhavr2013,pluhavr2014,kottos2001,kottos2003,schanz2003,wang2024bound}), are excluded from the analysis (see blue lines). Furthermore, the excellent agreement observed between the theoretical predictions, the chaotic cavity simulations, and the microwave network simulations underscores the robustness of the diagrammatic approach.

In greater detail, Fig.~\ref{Fig2}(a1) and \ref{Fig2}(a2) represent scenarios under CCC ($m_{\rm in} + m_{\rm tar} = 1$), differing only in the asymmetry between the interrogating and targeted channels. Under this constraint, the bimodal statistics expected for ideal coupling are skewed toward smaller transmittance eigenvalues. However, the scattering process still retains a moderate likelihood of supporting open channels, with $\tau_{\rm max} \approx 1$. For $m_{\rm in} = m_{\rm tar} = 1/2$ and no absorption (Fig.~\ref{Fig2}(a1)), Eq.~(\ref{Eq2}) simplifies to an explicit solution, ${\cal P}(\tau) = \frac{1}{\pi} \frac{\Gamma (2 - \Gamma)}{\sqrt{\tau (1 - \tau)} \, (\Gamma^2 - 4 \Gamma \tau + 4 \tau)}$ (see Supplementary Section II.C). A noteworthy difference between Figs.~\ref{Fig2}(a1) and \ref{Fig2}(a2) is the dramatic suppression of ${\cal P}(\tau)$ near $\tau \sim 0$ when $m_{\rm in} \neq m_{\rm tar}$, accompanied by an enhancement elsewhere.
When CCC is broken ($m_{\rm in} + m_{\rm tar} < 1$), open channels are strongly suppressed, leading to a gap in the ${\cal P}(\tau)$ distribution and the breakdown of its bimodal structure, as shown in Fig.~\ref{Fig2}(a3).

Including Ohmic losses further accentuates the gap in ${\cal P}(\tau)$, even under CCC, as illustrated in Figs.~\ref{Fig2}(b1) and \ref{Fig2}(b2). Despite this, the bimodal structure observed in Figs.~\ref{Fig2}(a1-a2) remains. The emergence of the gap near $\tau_{\rm max} \sim 1$ can be qualitatively understood by considering that losses effectively introduce additional uncontrolled output channels. However, unlike perfectly coupled uncontrolled channels, absorption channels partially backscatter the outgoing waves into the cavity (see Supplementary Section II.A). Violation of CCC primarily affects open channels (i.e., those with $\tau \approx 1$), whereas absorption has a more uniform impact across all channels.
In the combined presence of broken CCC and absorption, the bimodal nature of ${\cal P}(\tau)$ is entirely lost, and the gap is further enlarged, as seen in Fig.~\ref{Fig2}(b3).

\section{Comparison with experimental measurements}

\begin{figure*}
\centering
\includegraphics[width=\linewidth]{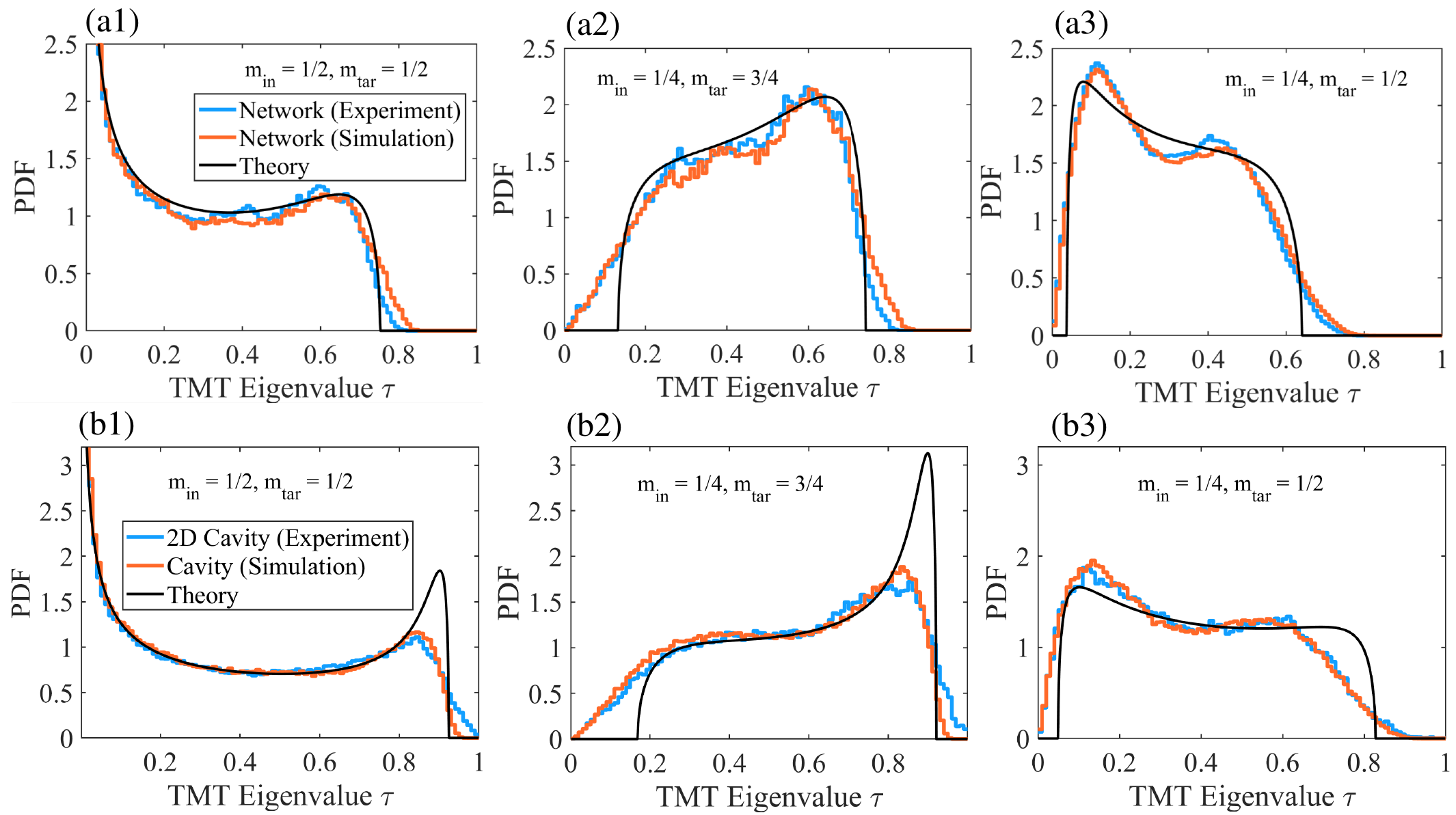}
\caption{\label{Fig3}
(a1)–(a3) Probability density function (PDF) of TMT eigenvalues of a fully connected network consisting of eight ``supervertices'' (see text), each connected to a TL with a coupling parameter $\Gamma \approx 0.97$. Various channel ratios $m_{\rm in}$ and $m_{\rm tar}$ are considered.  Experimental results (blue lines) and simulations (orange lines) are shown. Intrinsic cable losses are modeled with an imaginary part of the refractive index of $\approx 2\times 10^{-3}$, corresponding to absorption $A\approx 0.35$. The statistical analysis is performed over the frequency range [$1.5$ GHz, $4.5$ GHz].  Predictions from the diagrammatic approach are shown (black lines), with an absorption factor $a=A/(1-A/\Gamma) \approx 0.55$. (b1)–(b3) The same analysis as in (a1)–(a3), but for the 2D cavity shown in Fig. \ref{Fig1}(b) attached to eight TLs.  Experimental results (blue lines) 
and simulations (orange lines) are presented. Statistics for the experimental data are collected over the frequency range [$8$ GHz, $15$ GHz], while 
simulations use ensembles of random Gaussian matrices of dimensionality $100 \times 100$. The simulated system is characterized by a coupling 
parameter $\Gamma \approx 0.92$ and absorption $A \approx 0.1$. Predictions from the diagrammatic approach are shown (black lines), with an absorption factor $a=A/(1-A/\Gamma) \approx 0.11$.
}
\end{figure*}

Next, we compare the results of our theoretical model with experimental measurements conducted in complex networks, 2D chaotic cavities, and 3D-reverberation chambers. To facilitate the comparison, all cases involved $M=8$ channels. We first consider the microwave network consisting of 28 coaxial cables, as shown in Fig.~\ref{Fig1}(a). To achieve maximum all-to-all connectivity and ensure chaotic wave dynamics, we designed 8-port ``supervertices''—each consisting of a combination of six Tee-junctions. Each supervertex was coupled to a TL. In the frequency range [$1.5$ GHz, $4.5$ GHz] where the measurements were performed, the average coupling parameter was estimated as $\Gamma=1-|\langle S_{\alpha\alpha}\rangle|^2 \approx 0.97$ (see Methods and Supplementary Section I). From the measured scattering matrices, we extracted the TMT matrices $T$ and evaluated their eigenvalues $\tau$. Some typical TMT distributions ${\cal P}(\tau)$ for various $m_{\rm in}$, $m_{\rm tar}$ configurations are shown in Figs.~\ref{Fig3}(a1-a3) (blue lines). For comparison, we also report the results of simulations for the corresponding network with the same values of $M=8$ channels, coupling $\Gamma$, and absorption $A\approx 0.35$ as in the experiment (orange lines). The diagrammatic predictions, formally derived in the limit $M_{\rm in}, M_{\rm tar} \gg 1$, for $a=A/(1-A/\Gamma)\approx 0.55$ are also shown (black lines) in the same figures. The good agreement reveals the robustness of the analytical theory even for a moderate number of channels. 

The same experimental analysis was carried out for the 2D chaotic cavity shown in Fig.~\ref{Fig1}(b). Statistics (blue lines in Figs.~\ref{Fig3}(b1-b3)) were generated from scattering measurements in the frequency range [$8$ GHz, $15$ GHz]. Simulation results for the cavity, based on random matrix modeling (see orange lines), indicate that the microscopic model parameters that best fit the experimental data are $(\gamma,\gamma')=(0.56, 0.003)$, corresponding to the ensemble-averaged macroscopic model parameters $(\Gamma, A) \approx (0.92, 0.1)$. These values have to be compared with the experimentally estimated parameters $(\Gamma, A)\approx (0.95, 0.1)$. In the simulations, the ensemble average was performed over realizations of $100\times 100$ random Gaussian matrices, while in the experiment the ensemble was constructed over different frequencies. In the same figure, we also report the predictions of the diagrammatic approach (black lines), using $a=A/(1-A/\Gamma)\approx 0.11$. Overall, good agreement is observed between the experimental, numerical, and theoretical results.

A prominent feature in all cases shown in Fig.~\ref{Fig3} is the suppression of open channels and the formation of a statistical gap, primarily due to absorption in both setups. The greater absorption in the network compared to the cavity is evident from the more pronounced contraction of the eigenvalue spectrum toward smaller $\tau$ values.  It is also worth highlighting the asymmetric case of Figs.~\ref{Fig3}(a2, b2), where the distribution ${\cal P}(\tau)$ peaks near the upper bound of the TMT eigenvalues. A comparison with the analogous case in Fig.~\ref{Fig2}(b2) reveals that the critical factor here is the increased coupling parameter $\Gamma$, which approaches perfect coupling. 

When strong incomplete channel control or significant losses arise, as in the 3D cavity shown in Fig.~\ref{Fig1}(c), the correlations in the TMT matrix $T$ are progressively lost. In such cases, we can show that the distribution of eigenvalues normalized by their mean, ${\cal P}(x = \tau / \langle \tau \rangle)$, derived using the diagrammatic approach, converges to the Marchenko-Pastur law for rectangular random matrices with uncorrelated Gaussian elements, 
${\cal P}(x) = \frac{\sqrt{(x^+ - x)(x - x^-)}}{2\pi (m_{\rm in}/m_{\rm tar}) x}$, where $x^{\pm} = (1 \pm \sqrt{m_{\rm in}/m_{\rm tar}})^2$~\cite{marchenko1967distribution,gradoni2020statistical}. As derived, $\langle \tau \rangle =  m_{\rm tar} \Gamma/(1+a/\Gamma)$ (see Supplementary Section II.D), implying that the largest and smallest accessible transmission coefficients are $\tau^{\pm} =\Gamma (\sqrt{m_{\rm in}} \pm \sqrt{m_{\rm tar}})^2/(1+a/\Gamma)$, 
covering a range $\tau^+ - \tau^- = 4\Gamma \sqrt{m_{\rm in} m_{\rm tar}}/(1+a/\Gamma)$.
These results have been confirmed by measurements in the 3D reverberation chamber (see Supplementary Section III). 

\section{Extreme TMT Bounds and Conclusions}

Our diagrammatic approach takes advantage of the complex nature of scattering processes in wave-chaotic environments to derive a universal description of the dependence of the TMT eigenvalue density on intrinsic losses, coupling strength between the scattering domain and the interrogating/targeted antennas, and incomplete channel control ---factors often present in realistic operational scenarios, such as indoor wireless communications. 

The main findings of our analysis are as follows.
For lossless systems, the bimodal eigenvalue statistics predicted for perfect and symmetric coupling ($\Gamma = 1$, $m_{\rm in}=m_{\rm tar}=1/2$) becomes increasingly skewed toward smaller TMT eigenvalues as the coupling parameter decreases. In scenarios where CCC is preserved ($m_{\rm in} + m_{\rm tar} = 1$), the TMT eigenvalue distribution extends up to unity. A statistical gap near the open channels forms only for very weak coupling $\Gamma$ in asymmetric channel scenarios ($m_{\rm in} \neq m_{\rm tar}$). This gap widens in cases of broken CCC, where the bimodal nature of the statistics is entirely suppressed.
In the presence of losses, the statistical gap becomes even more pronounced and is observed even under CCC. 
An important result of our analysis is the identification of conditions under which the TMT distribution develops a skewed bimodal shape, peaking around the maximum eigenvalues. This optimal TMT scenario occurs when $\Gamma \rightarrow 1$ and the number of injected and targeted channels are unequal. We note that, due to reciprocity, all our results are also valid for $m_{\rm in}>m_{\rm tar}$ (apart from the $M_{\rm in}-M_{\rm tar}$ zero TMT eigenvalues).

\begin{figure}
\centering
\includegraphics[width=\linewidth]{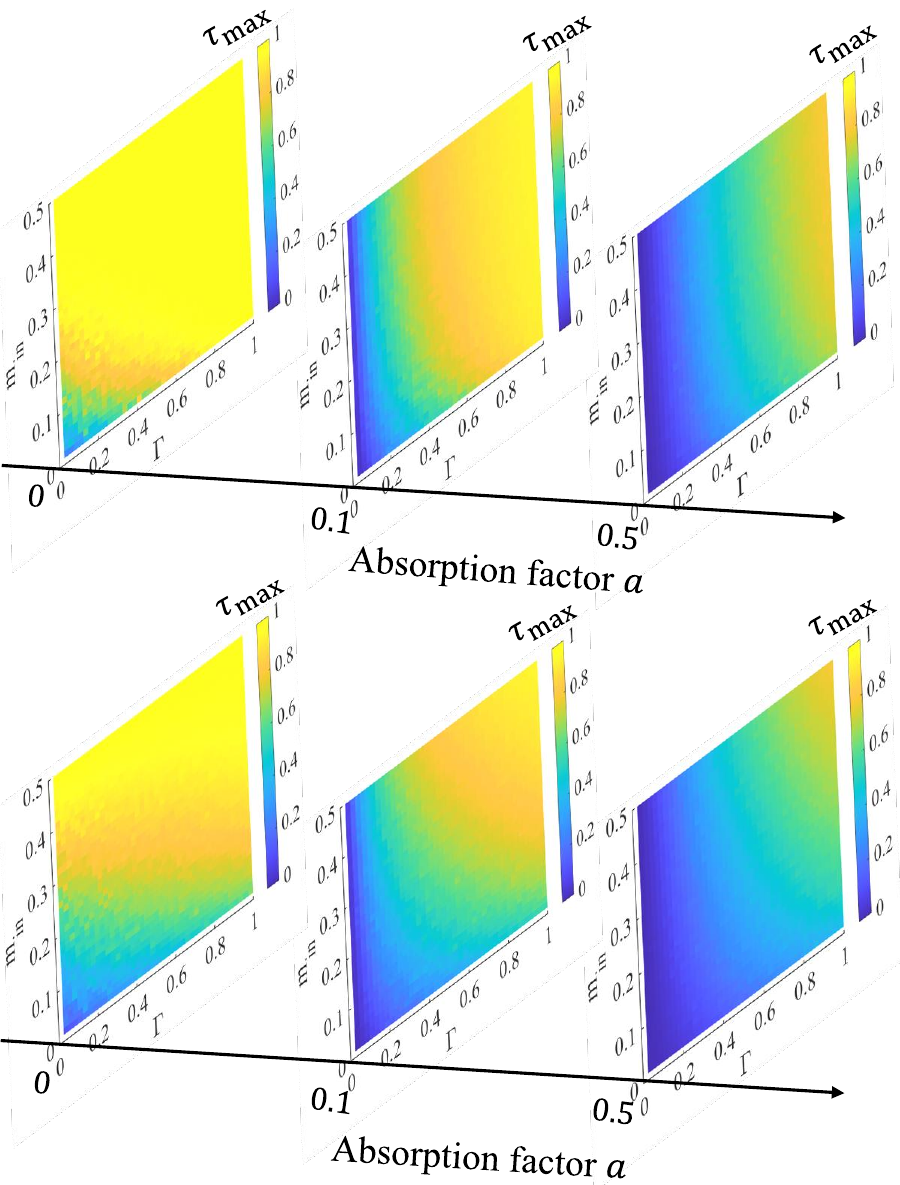}
\caption{ \label{Fig4}
 Maximum transmission $\tau_{\rm max}$ as a function of the input channel ratio $m_{in}$ and the coupling parameter $\Gamma$ for different absorption factors: $a=0$, $0.1$, and $0.5$ (left to right). The upper row corresponds to complementary channel constraint scenarios ($m_{\rm in} + m_{\rm tar} = 1$), while the lower row presents 
incomplete channel constraint scenarios with a fixed output channel ratio $m_{\rm out} = 0.5$. The simulation results shown here are based on a chaotic cavity modeled by an effective random Gaussian Hamiltonian with $N = 600$ modes and $M = 160$ TLs, and are in perfect agreement with the results of the diagrammatic approach (see Methods).
}
\end{figure}

The extreme TMT eigenvalues provide a direct estimate of the efficiency of CWS schemes. These extreme eigenvalues are predicted by our diagrammatic theory as the solutions of an explicit set of analytical equations (see Methods and Supplementary Section II.E).
Figure~\ref{Fig4} summarizes the dependence of the upper bound $\tau_{\rm max}$ on the absorption parameter $a$, the coupling parameter $\Gamma$, and the degree of channel control determined by $m_{\rm in}$ and $m_{\rm tar}$.  
The top row represents the scenario of CCC, where the maximum eigenvalue $\tau_{\rm max}$ achieves near-perfect TMT for $\Gamma \approx 1$, even as losses increase. Additionally, symmetric channel control ($m_{\rm in} \rightarrow m_{\rm tar}$) enhances the perfect TMT scenario.  Remarkably, we predict that any non-absorbing complex cavity will exhibit reflectionless states---as studied in Ref.~\cite{sweeney2020theory,sol2023reflectionless,jiang2024coherent} and defined as states with $\tau_{\rm max} = 1$ under CCC---irrespective of the fraction $m_{\rm in}$ of injected channels and for almost any coupling strength $\Gamma$. The bottom row of Fig.~\ref{Fig4} shows the scenario of non-complementary channel configurations ($m_{\rm in} + m_{\rm tar} <1$, with $m_{\rm tar}=0.5$). Here, perfect TMT is achieved when $\Gamma \rightarrow 1$, but incomplete channel control ($m_{\rm in} \rightarrow 0$) becomes detrimental even under ideal coupling ($\Gamma = 1$).  This analysis reveals that good TMT performance ($\tau_{\rm max}\gtrsim 0.8$) can still be preserved for moderate coupling ($\Gamma \gtrsim 0.5$) and absorption ($a \lesssim 0.1$).

Our theoretical predictions have been validated through comparisons with measurements across various microwave platforms. The universality of this formalism, driven by the intrinsic complex dynamics of wave-chaotic media, ensures its applicability to diverse settings, including optics, acoustics, and mechanical waves. Potential applications extend beyond indoor wireless communications to domains such as directed energy transfer, wireless power transfer, and more. Future work could explore extensions to scenarios where nonlinear interactions become dominant.

{\it Acknowledgements --} CW, JG, MR, and TK, acknowledge partial support from NSF-RINGS ECCS (2148318) and Simons
Foundation (SFI-MPS-EWP-00008530-08). AG acknowledges support from the program ``Investissements d’Avenir" launched by the French Government. MD acknowledges partial support by the European Union through European Regional Development Fund (ERDF), the Ministry of Higher Education and Research, CNRS, Brittany region, Conseils Départementaux d’Ille-et-Vilaine and Côtes d’Armor, Rennes Métropole, and Lannion Trégor Communauté, through the CPER Project CyMoCod.
\section{Methods}

\subsection{Description of experimental microwave network}
The microwave network is formed by coaxial cables (bonds) with physical lengths between $10$ cm and $50$ cm that 
are incommensurate with one another. These cables are connected with one-another via $N=8$ 
supervertices. Each supervertex is characterized by its valency $v_n$ ($n=1\cdots N$) indicating 
the number of cables that emanate from it and are connected to other supervertices of the network. In 
our case, all supervertices have been constructed to have $v=7$ and were assembled using six Tee-junctions (see Supplementary Section I for details). 
Five of these Tee-junctions have one female connector while the sixth Tee-junction consists of all male 
connectors. This setup results in a fully connected chaotic microwave network. At each supervertex, we 
have attached one transmission line (TL) supporting a single propagating mode connected to one port of 
a VNA which is used to measure the scattering parameters. The $8\times 8$ 
scattering matrix has been measured via multiple measurements using a two-port VNA. Each measurement 
utilized the two channels attached to the VNA while the rest six channels were connected to $50$ Ohm loads.
 
The network was interrogated in the frequency range [$1.5$ GHz, $4.5$ GHz], where the $8\times 8$ scattering matrix that describes the scattering process at the supervertex has matrix elements that are approximately constant. Within this range, the average coupling parameter $\Gamma$ between the network and the leads is constant and takes the approximate value $\Gamma\approx 0.97$. The high value of $\Gamma$ is associated with strong internal interferences occurring when the 6 T-junctions are combined together to create the super-vertex (see  Supplementary Section I for more details).

The experimental set-up involves Ohmic losses occurring at the cables. The loss of the cables is encoded 
in the imaginary part of its refractive index, which could be obtained via a best fitting of the measured 
frequency-dependent transmission ($t(\omega) = e^{i{\omega \over c} (n_r+in_i)L}$) through a cable of a 
specific length. Best fitting analysis indicated that $n_i\approx 2\times 10^{-3}$ while $n_r\approx 1.212$.

\subsection{Description of the experimental microwave cavities}
An image of the quasi-two-dimensional cavity of dimensions 
L = W = 205.74 mm and h = 10.16 mm is shown in Fig. \ref{Fig1}(b). The cavity is two-dimensional as only a single vertical polarization 
can propagate within the frequency range of interest [$8$ GHz, $15$ GHz]. In random positions inside the cavity, we have placed 
metallic cylinders that act as scatterers. The cavity is interrogated with eight antennas. We measure the $8\times8$ 
scattering matrix using a VNA between eight antennas that are matched coax-to-waveguide transitions attached to the cavity. A statistical ensemble of TMT eigenvalues is constructed by measuring the scattering matrix at various frequencies inside the operational frequency range with a frequency step of $0.28$ MHz. 

An image of the three-dimensional chaotic enclosure (reverberation chamber) of dimensions 
$1.75$ m$\times 1.5$ m$\times 2$ m is shown in Fig. \ref{Fig1}(c). The scattering matrix is measured 
between eight antennas that are commercial WiFi antennas 
(ANT-24G-HL90-SMA) matched at $2.4$ GHz. Two groups of four antennas are aligned and regularly spaced by $6.5$ cm, which is $\sim\lambda/2$ of the central frequency.  The orientation of the two groups is orthogonal to suppress direct paths. The reverberation chamber (RC) is equipped with two mechanical stirrers (a horizontal and a vertical one) that allow us to generate an ensemble of random configurations of the scattering matrix. An ensemble of $40$ random configurations is obtained from the rotation of the stirrers by steps of $3^o$. For better statistical processing of TMT eigenvalues, we have generated, for each cavity configuration, a number of scattering matrices corresponding to different frequencies in the range  [$2$ GHz, $3$ GHz].

\subsection{Scattering Theory for Networks}

Microwave networks, consisting of  $n=1,...,N$ vertices, are prototype systems that have been used successfully for the study 
of the universal properties of wave chaotic systems. Two vertices $n,m$ are coupled together via coaxial cables (bonds) of length 
$l_{nm}$. In the studies of wave-chaos, it is typically assumed that the bond-lengths are incommensurate with one-another \cite{kottos2003}. The 
position $x_{nm}$ on a $b\equiv(n,m)$ bond is defined to be $x_{nm}=0$ $(l_{nm})$ on vertex $n$ ($m$). The field $\psi_{b}(x_{nm})$ on 
each bond satisfies the Helmholtz equation
\begin{align} 
\frac{d^2}{dx_{nm}^2}\psi_{b}(x_{nm})+k^2\psi_{b}(x_{nm})=0,
\label{Helm}
\end{align}
where $k=\omega n/c_0$ is the wave number, $\omega$ is the angular frequency, $c_0$ is the speed of light in vacuum, $n=n_r+in_i$ 
is the complex-valued relative refraction index with imaginary part $n_i$ indicating the losses of the coaxial cables. 
The solution of Eq.~(\ref{Helm}) is $\psi_{b}(x) = \phi_{n}\frac{\sin k(l_{b} - x)}{\sin kl_{b}} + \phi_{m} \frac{\sin kx}{\sin kl_{b}}$, where $\psi_b(0)=\phi_{n}$ and $\psi_b(l_b)=\phi_{m}$ are the values of the field at the vertices.

We turn the compact network to a scattering set-up by attaching transmission lines (TL) $\alpha=1,\cdots, M$ to $M\leq N$ vertices. 
The field at the $\alpha-$TL takes the form $\psi_{\alpha}(x)={\cal I}_\alpha e^{-ikx}+{\cal O}_{\alpha}e^{+ikx}$ for $x\ge 0$ where 
$x=0$ is the position of the vertex and ${\cal I}_\alpha, {\cal O}_{\alpha}$ indicate the incoming and outgoing wave amplitudes at the $\alpha-$th TL. 

The field and current continuity on each vertex $n$ could be combined to give the scattering matrix \cite{kottos2003}
\begin{align} \label{Eq:Graph-S-Matrix}
S(\omega) = -\hat{I} + 2iD^T\frac{1}{{\cal M}(\omega)+iDD^T}D.
\end{align}
where the $N\times N$ matrix ${\cal M}(\omega)$ encodes the topology of the network (connectivity and length of bonds) and has elements
\begin{eqnarray} \label{Mmatrix}
	{\cal M}_{nm}(\omega) =\left\{
	\begin{array}{ll}
		-\sum_{l\neq n}{\cal A}_{nl}\cot kl_{nl}, & \textrm{if } n = m\\
		{\cal A}_{nm}\csc kl_{nm}, &\textrm{if } n \neq m
	\end{array}\right.
\end{eqnarray}
where ${\cal A}_{nl}$ takes the values $1 (0)$ when two vertices $n,l$ are (not) connected. Finally, the $N\times M$ matrix 
$D$ describes the connection between the transmission lines (TL) $\alpha=1,\cdots, M$ and the specific vertices where the TLs 
are attached. It has matrix elements $D_{n\alpha}=1$ if the $\alpha$th TL is attached to vertex $n$ and zero otherwise.\\

\subsection{Analytical solutions for the eigenvalue PDF}
 
The theoretical curves for the TMT eigenvalue distributions in Figs.~2 and~3 have been obtained by solving the equations $F_\text{in}(z) = 0$ and $F_\text{tar}(z) = 0$ for the self-energy components $\Sigma_\text{in}$ and $\Sigma_\text{tar}$, where
\begin{widetext}
\begin{align}
F_\text{in}(z) &= \frac{\Sigma_\text{in}}{1 - \Sigma_\text{in} \Sigma_\text{tar}} - \frac{a \Sigma_\text{in}}{(1 - \Sigma_\text{in} \Sigma_\text{tar})^2} - 
\frac{1 - \Gamma}{1 - (1 - \Gamma)\Sigma_\text{in} \Sigma_\text{tar}}
\left[\left(1 + \frac{\alpha m_\text{in}}{1 - \alpha} + \frac{\beta m_\text{tar}}{1 - \beta}\right)\Sigma_\text{in} + 
\frac{\Gamma}{\sqrt{z}(1 - \Gamma)}\frac{m_\text{in}}{1 - \beta}\right],
\label{EqFin}
\\
F_\text{tar}(z) &= \frac{\Sigma_\text{tar}}{1 - \Sigma_\text{in} \Sigma_\text{tar}} - \frac{a \Sigma_\text{tar}}{(1 - \Sigma_\text{in} \Sigma_\text{tar})^2} - 
\frac{1 - \Gamma}{1 - (1 - \Gamma)\Sigma_\text{in} \Sigma_\text{tar}}
\left[\left(1 + \frac{\alpha m_\text{in}}{1 - \alpha} + \frac{\beta m_\text{tar}}{1 - \beta}\right)\Sigma_\text{tar} + 
\frac{\Gamma}{\sqrt{z}(1 - \Gamma)}\frac{m_\text{tar}}{1 - \beta}\right],
\label{EqFtar}
\end{align}
\end{widetext}
with $\alpha = \frac{\Gamma \Sigma_\text{tar} / \sqrt{z}}{1 - (1 - \Gamma)\Sigma_\text{in} \Sigma_\text{tar}}$, $\beta = \frac{\Gamma \Sigma_\text{in} / \sqrt{z}}{1 - (1 - \Gamma)\Sigma_\text{in} \Sigma_\text{tar}}$, and inserting their values into Eq.~(2) of the main text. 

In addition, the upper bound $\tau_\text{max}$ shown in Fig.~4 can be obtained by solving the system of equations composed of $F_\text{in}(\tau_\text{max}) = 0$, $F_\text{tar}(\tau_\text{max}) = 0$, and $\partial_{\Sigma_{\rm in}} F_\text{in}(\tau_\text{max}) \partial_{\Sigma_{\rm tar}} F_\text{tar}(\tau_\text{max}) = 
\partial_{\Sigma_{\rm tar}} F_\text{in}(\tau_\text{max}) \partial_{\Sigma_{\rm in}} F_\text{tar}(\tau_\text{max})$.
These results are demonstrated in Supplementary Section II.

\bibliography{TTM_Statistics_paper2}
\end{document}


\title{Supplementary Material for ``Optimal Targeted Mode Transport in Complex Wave Environments: A Universal Statistical Framework"}

\author{Cheng-Zhen Wang}
\altaffiliation{These authors contributed equally to this work}
\affiliation{Wave Transport in Complex Systems Lab, Department of Physics, Wesleyan University, Middletown, CT-06459, USA}
\author{John Guillamon}
\altaffiliation{These authors contributed equally to this work}
\affiliation{Wave Transport in Complex Systems Lab, Department of Physics, Wesleyan University, Middletown, CT-06459, USA}
\author{Ulrich Kuhl}
\affiliation{Wave Transport in Complex Systems Lab, Department of Physics, Wesleyan University, Middletown, CT-06459, USA}
\affiliation{Université Côte d’Azur, CNRS, Institut de Physique de Nice (INPHYNI), 06200, Nice, France}
\author{Matthieu Davy}
\affiliation{Université de Rennes, CNRS, IETR- UMR 6164; F-35000 Rennes, France}
\author{Mattis Reisner}
\affiliation{Wave Transport in Complex Systems Lab, Department of Physics, Wesleyan University, Middletown, CT-06459, USA}
\author{Arthur Goetschy}
\email[]{ arthur.goetschy@espci.psl.eu}
\affiliation{ESPCI Paris, PSL University, CNRS, Institut Langevin, Paris, France}
\author{Tsampikos Kottos}
\email[]{tkottos@wesleyan.edu}
\affiliation{Wave Transport in Complex Systems Lab, Department of Physics, Wesleyan University, Middletown, CT-06459, USA}

\maketitle

\section{Experimental characterization and theoretical modeling of the super-vertices in networks}

The scattering process at each vertex of the network is characterized by a scattering matrix $\sigma$ which, in the absence of losses, is unitary. When a vertex is considered point-like, the implementation of field continuity and current conservation leads to the so-called Neumann vertex scattering matrix $\sigma^{(N)}_{nm}=-\delta_{nm}+{2\over v_n}$ where $v_n$ is the valency of the $n-$th vertex, indicating the number $m$ of other vertices of the network that are connected with it \cite{kottos2003}. In case that this vertex is coupled to a TL the above expression has to be modified by changing $v_n\rightarrow v_n+1$ in order to take into account the TL. An experimental realization of a Neumann vertex is provided by standard three-prong Tee-junctions where $v=3$ in the absence of TL, and $v=2$ when a TL is also attached to the vertex. 

The simplicity of the vertex Neumann scattering matrix allows us to evaluate immediately the coupling parameter which takes the frequency independent form $\Gamma_\alpha\equiv 1-|\langle \sigma^{(N)}_{\alpha,\alpha}\rangle|^2 =1 -\left(1-{2\over v_\alpha +1}\right)^2$.

In the case of super-vertices, however, the vertex scattering matrix is not (necessarily) modeled by a Neumann matrix. Instead, internal wave interferences occurring among the various Tee-junction components that form the super-vertex, lead to a rapidly fluctuating vertex scattering matrix with respect to frequency sweeping. As a result, the $\Gamma$ parameter has also strong frequency dependence. Additionally, losses in the super-junction might further complicate the scattering process.

\begin{figure*}[b]
\centering
\includegraphics[width=\linewidth]{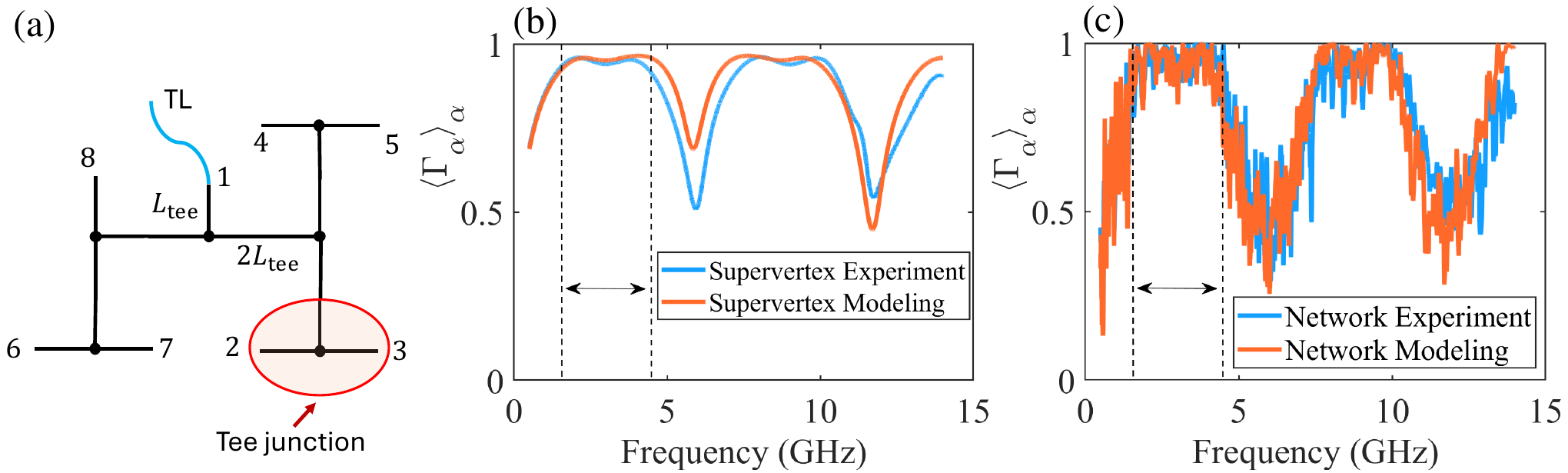}
\caption{ \label{FigS4} (a) Schematics of the eight-port supervertex consisting of six Tee-junctions coupled together as indicated in the figure. The TL (blue line) is indicatively coupled to one of the eight ports of the supervertex. (b) Coupling parameter $\langle\Gamma_\alpha\rangle_\alpha$ versus frequency in a broad frequency range for a single supervertex. Blue line correspond to the experimental data (averaged over the eight channels) while the orangle line are the results of our theoretical modeling. (c) The same as in (b) but now for the case of a network consisting of eight supervertices that are all coupled together.
}
\end{figure*}

We have modeled this behavior theoretically by explicitly incorporating in the supervertex scattering modeling, six Tee-junctions coupled together to create an eight-port supervertex, see Fig.~\ref{FigS4}(a). Internal losses within the supervertex has been modeled by an imaginary part in the refractive index, $n_i\approx 2\times 10^{-3}$. In Fig.~\ref{FigS4}(b) we report the results of our modeling for the coupling parameter $\langle\Gamma_\alpha\rangle_\alpha$ (orange line) against the measured coupling parameter for the eight-port supervertex (blue line). The averaging $\langle \cdots\rangle_\alpha$ is among the eight channels of the supervertex and the results are plotted versus frequency. The good agreement between theory and experiment indicate that our modeling captures successfully the scattering process at the supervertices.

The results reported in Fig.~\ref{FigS4}(b) indicate a strong frequency dependence of the coupling parameter. At the same time, we identify frequency domains where $\langle\Gamma_\alpha\rangle_\alpha$ is approximately constant. The same overall behavior of $\langle\Gamma_\alpha\rangle_\alpha$ versus frequency, appears also for the case of the fully connected network of eight supervertices, see Fig.~\ref{FigS4}(c). We have, therefore, chosen for the analysis of the probability distribution of the TMT eigenvalues, the frequency range [$1.5$ GHz, $4.5$ GHz]  where the coupling parameter is approximately constant with a mean value $\Gamma\approx 0.97$.

\newpage

\section{Theoretical predictions}

\subsection{Definition of the model}

Consider a lossless scattering setup, such as a chaotic cavity, involving $M$ physical channels characterized by a coupling parameter $\Gamma$. The transport properties of such a system are described by an $M$-dimensional scattering matrix $S$. When inherent (Ohmic) losses are introduced, the transport can instead be described by an effective scattering matrix $S_{\rm eff}$ of enlarged dimension $\tilde{M} = M + M'$. In this framework, the additional $M' \gg M$ channels account for the Ohmic losses and are characterized by a coupling parameter $\Gamma' \ll \Gamma$~\cite{brouwer97_prb,Fyodorov_2005}.

Furthermore, assuming all channels of the complex system are equivalent except for their coupling, we model the total system as comprising a perfectly coupled and lossless subsystem associated with a $\tilde{M} \times \tilde{M}$ unitary matrix $S_0$, attached to a barrier that encapsulates the effects of imperfect coupling. The $2\tilde{M} \times 2\tilde{M}$ scattering matrix of the barrier is expressed as:
\be
S_1=\begin{pmatrix}
r &  t'  \\
t & r' 
\end{pmatrix}, 
\ee
Without loss of generality, we express the transmission and reflection matrices of the barrier in a basis where they are diagonal. These matrices are given by:
\be
t=t'=\begin{pmatrix}
\sqrt{\Gamma}\mathds{1}_M &  0 \\
0 & \sqrt{\Gamma'}\mathds{1}_{M'} 
\end{pmatrix},
\label{EqTbarrier}
\ee
and
\be
r=-r'=\begin{pmatrix}
\sqrt{1-\Gamma}\mathds{1}_M &  0 \\
0 & \sqrt{1-\Gamma'}\mathds{1}_{M'}
\end{pmatrix}. 
\label{EqRbarrier}
\ee
By summing all scattering sequences contributing to the interaction between the system's barrier and the perfectly coupled subsystem, we obtain the following expression for the total scattering matrix:
\begin{align}
S_{\rm eff}&= r+ t'S_0t+t'S_0r'S_0t + t'(S_0r')^2S_0t+ \dots
\nonumber
\\
&=
r+ t' \frac{1}{\mathds{1}_{\tilde{M}} - S_0r'}S_0t.
\label{EqSeff}
\end{align}

Using the projectors $P_\text{in} = \mathds{1}_{M_\text{in}}$ and $P_\text{tar} = \mathds{1}_{M_\text{tar}}$, which act on the $M$-dimensional (physical) channel subspace with $M_\text{in}, M_\text{tar} \leq M$, and satisfy the orthogonality constraint $P_\text{in} \cdot P_\text{tar} = 0$, we obtain the $M_\text{tar} \times M_\text{in}$ scattering submatrix $\tilde{S}$:  
\begin{align}
\tilde{S} = P_\text{tar} S_\text{eff} P_\text{in} = P_\text{tar} t' \frac{1}{1 - S_0 r'} S_0 t P_\text{in},
\end{align}
from which we evaluate the $M_\text{in} \times M_\text{in}$ Hermitian TMT matrix $T = \tilde{S}^\dagger \tilde{S}$. In the following, we will focus on the evaluation of the resolvent:  
\begin{equation}
g_T(z) = \frac{1}{M_{\rm in}} \left\langle \text{Tr} \frac{1}{z - T} \right\rangle,
\label{EqGT}
\end{equation}
which provides access to the density of eigenvalues $\tau$ of the matrix $T$ via  
${\cal P}(\tau) = -\frac{1}{\pi} \lim_{\eta \to 0^+} \text{Im}[g_T(\tau + i\eta)]$.

\subsection{A special case: perfect coupling and no absorption}

In the simple case of perfect coupling ($\Gamma = 1$) and no absorption ($\Gamma' = 0$), the filtered scattering matrix can be expressed as $\tilde{S} = P_\text{tar} S_0 P_\text{in}$, where $S_0$ is an $M \times M$ matrix uniformly distributed within the unitary group. Since all scattering channels of the matrix  to be filtered are statistically equivalent in this scenario, we can use the filtered random matrix (FRM) theory~\cite{goetschy2013filtering}, which provides $g_T(z)$ as the solution of the implicit equation:  
\be
\label{SelfG}
N(z) g_{S_0^\dagger S_0} \left[\frac{N(z)^2}{D(z)}\right] = D(z).
\ee
Here, $N(z)$ and $D(z)$ are auxiliary functions defined as:  
\begin{align}
N(z) &= z \, m_\text{in} g_T(z) + 1 - m_\text{in}, \\
D(z) &= m_\text{in} g_T(z) \left[ z m_\text{in} g_T(z) + m_\text{tar} - m_\text{in} \right],
\end{align}
with $m_\text{in} = M_\text{in} / M$ and $m_\text{tar} = M_\text{tar} / M$. Since $S_0^\dagger S_0 = \mathds{1}_M$, its resolvent has a simple form: $g_{S_0^\dagger S_0}(z) = 1 / (z - 1)$. The solution of Eq.~\eqref{SelfG} is given by:  
\begin{align}
g_T(z) = \frac{1}{2 m_\text{in} z (1 - z)} \left[ m_\text{in} - m_\text{tar} + (1 - 2 m_\text{in}) z 
- \sqrt{(m_\text{in} - m_\text{tar})^2 + z (z - 2 m_\text{in} - 2 m_\text{tar} + 4 m_\text{in} m_\text{tar})} \right],
\label{EqSolGTFRM}
\end{align}
from which we deduce the eigenvalue density:  
\begin{equation}
\mathcal{P}(\tau) = \frac{1}{\pi} \frac{\sqrt{(\tau^+ - \tau)(\tau - \tau^-)}}{2 m_\text{in} \tau (1 - \tau)} 
+ \text{max}\left(1 - \frac{m_\text{tar}}{m_\text{in}}, 0\right)\delta(\tau),
\label{PDFPerfect}
\end{equation}
where the bounds $\tau^\pm$ are given by:  
\begin{align}
\tau^{\pm} = m_{\rm in} + m_{\rm tar} - 2 m_{\rm in} m_{\rm tar}
\pm 2 \sqrt{m_{\rm in} m_{\rm tar} (1 - m_{\rm in})(1 - m_{\rm tar})}.
\end{align}

We note, in particular, that under the complementary channel constraint (CCC) $m_{\rm in} + m_{\rm tar} = 1$, the upper bound simplifies to $\tau^{+} = 1$ for all $m_{\rm in}$. This indicates a non-zero probability of finding reflectionless states in the system. On the other hand, the lower bound $\tau^{-} = (1 - 2m_{\rm in})^2$ is never zero except in the symmetric case $m_{\rm in} = m_{\rm tar} = 1/2$.

\subsection{Self-consistent equations in the general case}

To evaluate Eq.~\eqref{EqGT}, we use a diagrammatic approach that takes advantages of the fact that $S_0$ is uniformly distributed within the unitary group \cite{brouwer96}. However, a direct series expansion of Eq.~\eqref{EqGT} in powers of $S_0$ is not suitable. A more convenient approach is to work in a duplicated space and introduce 
\be
\hat{S}=\begin{pmatrix}
S^\dagger &  0  \\
0 & S
\end{pmatrix}, \;\;
\hat{P}=\begin{pmatrix}
0 &  P_\text{tar}  \\
P_\text{in} & 0
\end{pmatrix}, 
\ee
as well as the matrix resolvent
\be
\hat{G}(z)=\begin{pmatrix}
0 & \hat{G}_\text{tar} \\
\hat{G}_\text{in} & 0
\end{pmatrix}, 
\ee
where ${\displaystyle\hat{G}_\text{in}(z) = \left\langle \frac{1}{z - \tilde{S}^\dagger \tilde{S}} \right\rangle}$ and ${\displaystyle \hat{G}_\text{tar}(z) = \left\langle \frac{1}{z - \tilde{S} \tilde{S}^\dagger} \right\rangle}$. Thus, $g_T(z) = \text{Tr}[\hat{G}_\text{in}(z)]/M_\text{in}$. The matrix resolvent can then be written as
\begin{align}
\hat{G}(z) = \left\langle \frac{1}{z - (\hat{P}\hat{S})^2} \hat{P} \right\rangle
= \frac{1}{2\sqrt{z}} \sum_{\pm} \left\langle \frac{1}{\sqrt{z} \pm \hat{P}\hat{S}} \hat{P} \right\rangle.
\end{align}
Noting that $\hat{P} \hat{S} \hat{P} = \hat{P} \hat{t} \left[\mathds{1}_{2\tilde{M}} - \hat{S}_0 \hat{r}'\right]^{-1} \hat{S}_0 \hat{t} \hat{P}$, where the following matrices have been introduced:
\be
\hat{S}_0 = \begin{pmatrix}
S_0^\dagger &  0  \\
0 & S_0
\end{pmatrix}, \;
\hat{t} = \begin{pmatrix}
t^\dagger &  0  \\
0 & t
\end{pmatrix}, \;
\hat{r}' = \begin{pmatrix}
r'^\dagger &  0  \\
0 & r'
\end{pmatrix},
\ee
simple algebraic manipulation allows the matrix resolvent to be expressed as
\begin{align}
\hat{G}(z) = \frac{\hat{P}}{2z} & \sum_{\pm} 
\left[ \mathds{1}_{2\tilde{M}} 
\pm \frac{1}{\sqrt{z}} \hat{t}(\hat{X}^\pm)^{-1} (\hat{G}^\pm - \hat{X}^\pm)(\hat{X}^\pm)^{-1} \hat{t} \hat{P} 
\right],
\end{align}
where $\hat{X}^\pm = \hat{r}' \pm \hat{t} \hat{P} \hat{t}/\sqrt{z}$ and
\be
\hat{G}^\pm = \left\langle \frac{1}{(\hat{X}^\pm)^{-1} - \hat{S}_0} \right\rangle = \frac{1}{(\hat{X}^\pm)^{-1} - \hat{\Sigma}^\pm}.
\label{EqGpm}
\ee
Hence, the diagrammatic evaluation of $\hat{G}_\text{in}(z)$ has been reduced to the evaluation of $\hat{G}^\pm$, which is technically less involved.

By combining the explicit expression of the matrix propagator, $\hat{X}^\pm = -\sqrt{\mathds{1}_{2\tilde{M}} - \hat{\Gamma}} \pm \hat{\Gamma}\hat{P}/\sqrt{z}$,
where $\hat{\Gamma} = \text{diag}(\Gamma \mathds{1}_{M}, \Gamma' \mathds{1}_{M'}, \Gamma \mathds{1}_{M}, \Gamma' \mathds{1}_{M'})$, with the fact -- demonstrated below -- that the self-energy introduced in Eq.~\eqref{EqGpm} has the following simple form:
\be
\hat{\Sigma}^\pm = \begin{pmatrix}
0 & \Sigma_\text{tar}^\pm \mathds{1}_{\tilde{M}} \\
\Sigma_\text{in}^\pm \mathds{1}_{\tilde{M}} & 0
\end{pmatrix}, 
\ee
we find that the resolvent we are seeking can be expressed in terms of the self-energy elements as
\be
g_T(z) = \frac{1}{z} \frac{1 - (1 - \Gamma)\Sigma_{\text{in}}^{+}\Sigma_{\text{tar}}^{+}}{1 - (1 - \Gamma)\Sigma_{\text{in}}^{+}\Sigma_{\text{tar}}^{+} - \Gamma \Sigma_{\text{tar}}^{+} / \sqrt{z}}.
\label{EqSolGT}
\ee
To derive this expression, we used the orthogonality condition $P_\text{in} \cdot P_\text{tar} = 0$ and the relation $\hat{\Sigma}^- = -\hat{\Sigma}^+$.
Note that the superscripts in $\Sigma_{\text{in}}^{+}$ and $\Sigma_{\text{tar}}^{+}$ are omitted in Eq.~(2) of the main text for simplicity of notation.

Finally, we evaluate the self-energy matrix $\hat{\Sigma}^\pm$ defined in Eq.~\eqref{EqGpm} using a diagrammatic approach. By definition, its elements include all irreducible diagrams in the expansion of $(\hat{X}^\pm)^{-1}\hat{G}^\pm(\hat{X}^\pm)^{-1} = \hat{S}_0 + \hat{S}_0 \hat{X}^\pm \hat{S}_0 + \dots$. The average of the product of $\hat{S}_0$ matrix elements is given by~\cite{mello90, brouwer96}
\begin{align}
&\overline{(S_{0,a_1b_1} \dots S_{0,a_mb_m})(S_{0,\alpha_1\beta_1}^* \dots S^*_{0,\alpha_m\beta_m})} = 
\sum_{P,P'} V_{PP'} \prod_{j=1}^m \delta_{a_j\alpha_{P(j)}} \delta_{b_j\beta_{P'(j)}},
\label{EqUnitary}
\end{align}
where $P, P'$ are permutations of $\{1, \dots, m\}$, and the weights $V_{PP'}$ depend only on the cycle structure of $P^{-1}P'$: $V_{PP'} = V_{c_1 \dots c_k}$, with $c_j$ the lengths of disjoint cyclic permutations in $P^{-1}P'$ ($\sum_{j=1}^k c_j = m$). In the limit $\tilde{M} \gg 1$, it is known that~\cite{brouwer96}
\be
V_{c_1 \dots c_k} = \prod_{j=1}^k V_{c_j},\;  \text{with} \; V_n = \frac{(-1)^{n-1}(2n-2)!}{n(n-1)!^2} \frac{1}{\tilde{M}^{2n-1}}.
\ee

Among all possible contractions in Eq.~\eqref{EqUnitary}, only cycles corresponding to planar diagrams contribute significantly to the self-energy for $M_\text{in}, M_\text{tar} \gg 1$. These diagrams correspond to non-disjoint cyclic permutations, with $V_{PP'} \simeq V_n$. Summing over all planar diagrams, we find
\begin{align}
\Sigma_\text{in}^\pm &= V_1 \text{Tr} \hat{X}^\pm_{21} + V_2 (\text{Tr} \hat{X}^\pm_{21})^2 \text{Tr} \hat{X}^\pm_{12} + \dots
\nonumber
\\
&= \sum_{n=1}^\infty V_n (\text{Tr} \hat{G}_\text{in}^\pm )^n (\text{Tr} \hat{G}_\text{tar}^\pm )^{n-1}
\nonumber
\\
&= \text{Tr} \hat{G}_\text{in}^\pm \, H\left(\text{Tr} \hat{G}_\text{in}^\pm \text{Tr} \hat{G}_\text{tar}^\pm\right),
\label{EqSelf1}
\end{align}
where $\hat{G}_\text{in}^\pm$ and $\hat{G}_\text{tar}^\pm$ are the non-diagonal blocks of $\hat{G}^\pm$, and 
\begin{align}
H(z) &= \frac{1}{\tilde{M}} \sum_{n=0}^\infty \left(\frac{-z}{\tilde{M}^2}\right)^n \frac{(2n)!}{n!(n+1)!} 
\nonumber
\\
&= \frac{\sqrt{\tilde{M}^2 + 4z} - \tilde{M}}{2z}
\end{align}
is the generating function of the Catalan numbers. Similarly, we find
\be
\Sigma_\text{tar}^\pm = \text{Tr} \hat{G}_\text{tar}^\pm \, H\left(\text{Tr} \hat{G}_\text{in}^\pm \text{Tr} \hat{G}_\text{tar}^\pm\right).
\label{EqSelf2}
\ee
Since $\hat{S}_0$ contains only diagonal blocks, the diagonal blocks of $\hat{\Sigma}^\pm$ arise from non-planar diagrams, which can be neglected for $M_\text{in}, M_\text{tar} \gg 1$: $\hat{\Sigma}^\pm_{11} \simeq \hat{\Sigma}^\pm_{22} \simeq 0$.

By inverting the relations~\eqref{EqSelf1} and~\eqref{EqSelf2}, we express $\text{Tr} \hat{G}_\text{in}^\pm$ and $\text{Tr} \hat{G}_\text{tar}^\pm$ in terms of the self-energy elements:
\begin{align}
\text{Tr} \hat{G}_\text{in}^\pm &= \frac{\tilde{M} \Sigma_\text{in}^\pm}{1-\Sigma_\text{in}^\pm \Sigma_\text{tar}^\pm},
\nonumber
\\
\text{Tr} \hat{G}_\text{tar}^\pm &= \frac{\tilde{M} \Sigma_\text{tar}^\pm}{1-\Sigma_\text{in}^\pm \Sigma_\text{tar}^\pm}.
\end{align}
Equating these expressions to the traces of the diagonal blocks of Eq.~\eqref{EqGpm}, we find that the self-energy terms satisfy two coupled non-linear equations:
\be
F^\pm_\text{in}(z) = 0 \quad \text{and} \quad F^\pm_\text{tar}(z) = 0,
\label{EqCoupledNonLinear}
\ee
defined as
\begin{align}
F^\pm_\text{in}(z) &= \frac{\Sigma_\text{in}^\pm}{1-\Sigma_\text{in}^\pm \Sigma_\text{tar}^\pm} - \frac{a\Sigma_\text{in}^\pm}{(1-\Sigma_\text{in}^\pm \Sigma_\text{tar}^\pm)^2} - \frac{1-\Gamma}{1-(1-\Gamma)\Sigma_\text{in}^\pm \Sigma_\text{tar}^\pm}
\left[\left(1+\frac{\alpha_\pm m_\text{in}}{1-\alpha_\pm}+\frac{\beta_\pm m_\text{tar}}{1-\beta_\pm}\right)\Sigma_\text{in}^\pm 
\pm \frac{\Gamma}{\sqrt{z}(1-\Gamma)}\frac{m_\text{in}}{1-\beta_\pm} \right],
\label{EqFin}
\\
F^\pm_\text{tar}(z) &= \frac{\Sigma_\text{tar}^\pm}{1-\Sigma_\text{in}^\pm \Sigma_\text{tar}^\pm} - \frac{a\Sigma_\text{tar}^\pm}{(1-\Sigma_\text{in}^\pm \Sigma_\text{tar}^\pm)^2} - 
\frac{1-\Gamma}{1-(1-\Gamma)\Sigma_\text{in}^\pm \Sigma_\text{tar}^\pm}
\left[\left(1+\frac{\alpha_\pm m_\text{in}}{1-\alpha_\pm}+\frac{\beta_\pm m_\text{tar}}{1-\beta_\pm}\right)\Sigma_\text{tar}^\pm 
\pm \frac{\Gamma}{\sqrt{z}(1-\Gamma)}\frac{m_\text{tar}}{1-\beta_\pm} \right],
\label{EqFtar}
\end{align}
where $\alpha_\pm$ and $\beta_\pm$ are auxiliary functions:
\begin{align}
\alpha_\pm &= \frac{\pm \Gamma \Sigma_\text{tar}^\pm / \sqrt{z}}{1-(1-\Gamma)\Sigma_\text{in}^\pm \Sigma_\text{tar}^\pm},
\\
\beta_\pm &= \frac{\pm \Gamma \Sigma_\text{in}^\pm / \sqrt{z}}{1-(1-\Gamma)\Sigma_\text{in}^\pm \Sigma_\text{tar}^\pm},
\end{align}
and $a$ is the macroscopic absorption parameter in the limit $\Gamma' \to 0$ and $M' \to \infty$, with $M'\Gamma'/M$ finite:
\be
a = \frac{M'\Gamma'}{M} \frac{1-\Sigma_\text{in}^\pm \Sigma_\text{tar}^\pm}{1-(1-\Gamma')\Sigma_\text{in}^\pm \Sigma_\text{tar}^\pm} \xrightarrow[\Gamma'\to 0]{M'\to\infty} \frac{M'\Gamma'}{M}.
\ee
In the case of a lossy cavity supporting a very large number $N$ of modes and characterized by the loss rate $\gamma'$, we have $M' \simeq N$ and $\Gamma' = 1 - \frac{(1 - \gamma')^2}{(1 + \gamma')^2} \simeq 4\gamma'$ ($\gamma'\ll 1$), so that the absorption parameter can also be expressed as $a = 4(N / M)\gamma'$. As a result, the $M'$ channels do not introduce any additional fluctuations in the scattering properties of the cavity.
The resolvent of the TMT operator is obtained by substituting the solutions $\Sigma_\text{in}^+$ and $\Sigma_\text{tar}^+$ from Eq.~\eqref{EqCoupledNonLinear} into Eq.~\eqref{EqSolGT}. Note that the superscripts in $F_{\text{in}}^{+}$ and $F_{\text{tar}}^{+}$ are omitted in the Methods section of the main text for simplicity of notation.

We point out that Eq.~\eqref{EqCoupledNonLinear} has explicit analytical solutions only in some special cases. When there is no absorption ($a=0$) and the coupling is optimal ($\Gamma=1$), we find 
\begin{align}
\Sigma_\text{in}^\pm &= \pm\frac{-\sqrt{z}(z+m_\text{in}-m_\text{tar})+\sqrt{f(z,m_\text{in},m_\text{tar})}}{2(m_\text{tar}-1)z}, \\
\Sigma_\text{tar}^\pm &= \pm\frac{-\sqrt{z}(z+m_\text{tar}-m_\text{in})+\sqrt{f(z,m_\text{in},m_\text{tar})}}{2(m_\text{in}-1)z},
\end{align}
where $f(z,m_\text{in},m_\text{tar})= z^3+2(2m_\text{in}m_\text{tar}-m_\text{in}-m_\text{tar})z^2+(m_\text{in}-m_\text{tar})^2z$.
Combining these expressions with Eq.~\eqref{EqSolGT}, we recover the solution~\eqref{EqSolGTFRM}. 
On the other hand, when $\Gamma \neq 1$, an explicit solution can still be found in the special case of balanced CCC ($m_\text{in}=m_\text{tar}=1/2$) without absorption. Then, the self-energy terms are given by $\Sigma_\text{in}^+ = \Sigma_\text{tar}^+ = \sqrt{z} - \sqrt{z-1}$, 
for all $\Gamma$. The resolvent~\eqref{EqSolGT} becomes
\be
g_T(z) = \frac{1+(1-\Gamma)\left[\sqrt{z(z-1)} -z\right]-\Gamma/2}{z(1-\Gamma)\left[1+\sqrt{z(z-1)}-z\right]+\sqrt{z(z-1)}\Gamma/2},
\ee
which gives the bimodal eigenvalue density
\be
\mathcal{P}(\tau) = \frac{1}{\pi} \frac{\Gamma (2 - \Gamma)}{\sqrt{\tau (1 - \tau)} \, (\Gamma^2 - 4 \Gamma \tau + 4 \tau)},
\ee
in agreement with the results of Ref.~\cite{brouwer96}.

\subsection{Mean transmittance and mean absorption}

\begin{figure*}
\centering
\includegraphics[width=0.9\linewidth]{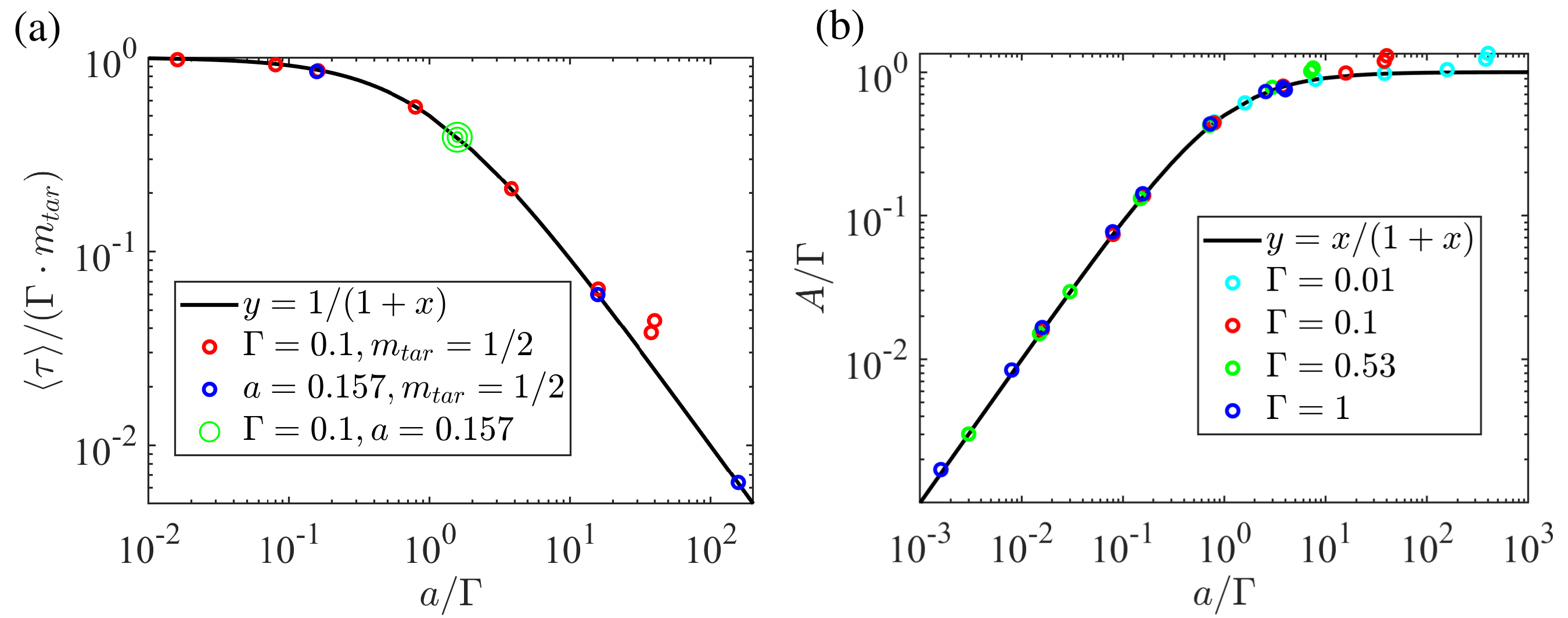}
\caption{ \label{FigS3} 
(a) Rescaled mean TMT eigenvalue $\langle \tau \rangle / m_{\rm tar}\Gamma$ versus the rescaled absorption factor $a/\Gamma$ for various combinations of $\Gamma$, $a$, and $m_{\rm tar}$ parameters. The legend indicates the parameters kept constant while varying $a$ (red circles), $\Gamma$ (blue circles), and $m_{\rm tar}$ (green circles of varying sizes, with smaller to larger circles corresponding to $M_{\rm in} = 10, 10, 30$ and $M_{\rm tar} = 40, 70, 50$, respectively). The black line represents the theoretical prediction $\langle \tau \rangle / m_{\rm tar}\Gamma = \frac{1}{1 + a/\Gamma}$. 
(b) Relationship between total absorption $A$ and the absorption factor $a$ for different coupling parameters $\Gamma$ (indicated by circles of different colors). All curves collapse onto a single universal curve when the $(x, y)$ axes are rescaled as $(a/\Gamma, A/\Gamma)$. The black line represents the theoretical prediction $A/\Gamma = \frac{a/\Gamma}{1 + a/\Gamma}$. 
In both subfigures, the symbols correspond to results of cavity simulations using a random matrix theory model for matrices with system size $N = 300$ and $M = 80$ channels.
}
\end{figure*}

The mean TMT eigenvalue can be derived from the distribution $\mathcal{P}(\tau)$. However, extracting its analytical expression directly from the self-consistent equations is tedious. Instead, we compute the mean directly using
\begin{widetext}
\begin{align}
\left< \tau \right> = \frac{1}{M_\text{in}}\left<\text{Tr} \left(\tilde{S}^\dagger\tilde{S} \right) \right>
&=\frac{1}{M_\text{in}}
\left<
\text{Tr} \left(P_\text{in} t^\dagger S_0^\dagger [1 - r'^\dagger S_0^\dagger]^{-1} t^{'\dagger} P_\text{tar} t' [1 - S_0 r']^{-1} S_0 t P_\text{in}
\right)
\right>
\nonumber
\\
&=\frac{1}{M_\text{in}}\sum_{n=0}^\infty
\left<
\text{Tr} \left(P_\text{in} t^\dagger S_0^\dagger  (r'^\dagger S_0^\dagger)^n t^{'\dagger} P_\text{tar} t' (S_0 r')^n S_0 t P_\text{in}
\right)
\right>.
\label{EqMean1}
\end{align}
\end{widetext}
In the last equality, we used the fact that a product involving random unitary matrices $S_0$ and $S_0^\dagger$ is non-zero only if it contains the same number of $S_0$ and $S_0^\dagger$ matrices. Using Eq.~\eqref{EqUnitary}, the leading planar diagrams in Eq.~\eqref{EqMean1} are rainbow diagrams of weight $V_1^{n+1} = 1/\tilde{M}^{n+1}$. 
This result is identical to what would be obtained when treating $S_0$ as a Gaussian random matrix. Equation~\eqref{EqMean1} then simplifies to
\begin{align}
\left< \tau \right>&= \frac{\text{Tr}( t P_\text{in} t^\dagger )\text{Tr}(t^{'\dagger} P_\text{tar} t')}{M_\text{in} \tilde{M}}\sum_{n=0}^\infty\left[\frac{\text{Tr}(r^{'\dagger}r')}{\tilde{M}}\right]^n
\nonumber
\\
&=\frac{\Gamma^2 M_\text{tar}}{\tilde{M}-\text{Tr}(r^{'\dagger}r')}
\nonumber
\\
&=\frac{m_\text{tar} \Gamma}{1+a/\Gamma},
\label{EqMeanTau}
\end{align}
where we used $\text{Tr}(r^{'\dagger}r') = M(1-\Gamma) + M'(1-\Gamma')$.

Similarly, the mean total absorption in the system is given by
\begin{align}
A&=1-\frac{1}{M}\left< \text{Tr}\left(S^\dagger S\right)\right>
\nonumber
\\
&=1-\frac{1}{M}\left< \text{Tr}\left(PS_\text{eff}^\dagger P S_\text{eff}P\right)\right>
\nonumber
\\
&=\frac{1}{M}\left< \text{Tr}\left(PS_\text{eff}^\dagger Q S_\text{eff}P\right)\right>,
\end{align}
where we introduced the projectors $P = \mathds{1}_{\text{M}}$ and $Q = \mathds{1}_{\tilde{\text{M}}} - P = \mathds{1}_{\text{M'}}$, and used the fact that the effective scattering matrix $S_\text{eff}$ is unitary, unlike $S$. Combining this with the orthogonality condition $P \cdot Q = 0$ and the expression~\eqref{EqSeff}, we find
\begin{align}
A&=\frac{1}{M}\left< \text{Tr} \left[
P t^\dagger S_0^\dagger [1 - r'^\dagger S_0^\dagger]^{-1} t^{'\dagger} Q t' [1 - S_0 r']^{-1} S_0 t P
\right]
\right>.
\end{align}
This expression differs from Eq.~\eqref{EqMean1} only by the projectors. Selecting the rainbow diagrams, it simplifies to
\begin{align}
A&= \frac{\text{Tr}( t P t^\dagger )\text{Tr}(t^{'\dagger} Q t')}{M_\text{in} \tilde{M}}\sum_{n=0}^\infty\left[\frac{\text{Tr}(r^{'\dagger}r')}{\tilde{M}}\right]^n
\nonumber
\\
&=\frac{\Gamma \Gamma' M'}{\tilde{M}-\text{Tr}(r^{'\dagger}r')}
\nonumber
\\
&=\frac{a}{1+a/\Gamma}.
\label{EqMeanA}
\end{align}

In Fig.~\ref{FigS3}, we compare the theoretical predictions~\eqref{EqMeanTau} and~\eqref{EqMeanA} for the mean transmission and mean absorption with numerical simulations of transport through a chaotic absorbing cavity using random matrix modeling of the effective Hamiltonian. Excellent agreement is found over a wide range of the absorption-to-radiative loss ratio $a/\Gamma$.

\subsection{Equations for the edges of the eigenvalue spectrum}

The bounds $z^*$ of the TMT eigenvalue distribution are determined by the condition $\partial_z g_T(z^*) = \infty$. From Eq.~\eqref{EqSolGT}, this condition is equivalent to $\partial_z \Sigma^+_\text{in} \vert_{z^*} = \infty$, which in turn implies $\partial_z \Sigma^+_\text{tar} \vert_{z^*} = \infty$, given the symmetry $\Sigma^+_\text{in}(z, m_\text{in}, m_\text{tar}) = \Sigma^+_\text{tar}(z, m_\text{tar}, m_\text{in})$. This symmetry ensures that exchanging $m_\text{in}$ and $m_\text{tar}$ does not alter the support of the non-zero TMT eigenvalue distribution. Differentiating Eq.~\eqref{EqCoupledNonLinear} and using the conditions $\partial_{\Sigma^+_\text{tar}} z \vert_{z^*} = \partial_{\Sigma^+_\text{in}} z \vert_{z^*} = 0$, we arrive at the equation 
\begin{align}
 \frac{\partial_{\Sigma^+_{\rm in}} F_\text{in}(z^*)}
 {\partial_{\Sigma^+_{\rm in}} F_\text{tar}(z^*)}
  = \frac{\partial_{\Sigma^+_{\rm tar}} F_\text{in}(z^*)}
 {\partial_{\Sigma^+_{\rm tar}} F_\text{tar}(z^*)},
 \label{EqEdge}
 \end{align}
 which can be explicitly expressed using Eqs.~\eqref{EqFin} and~\eqref{EqFtar}. The largest solution $z^*$ of the system of equations composed of $F_\text{in}(z^*) = 0$, $F_\text{tar}(z^*) = 0$, and Eq.~\eqref{EqEdge} defines the upper bound of the TMT distribution, while the smallest solution defines the lower bound.
\section{Marcenko-Pastur Law for 3D reverberation chambers}

\begin{figure*}[t]
\centering
\includegraphics[width=0.8\linewidth]{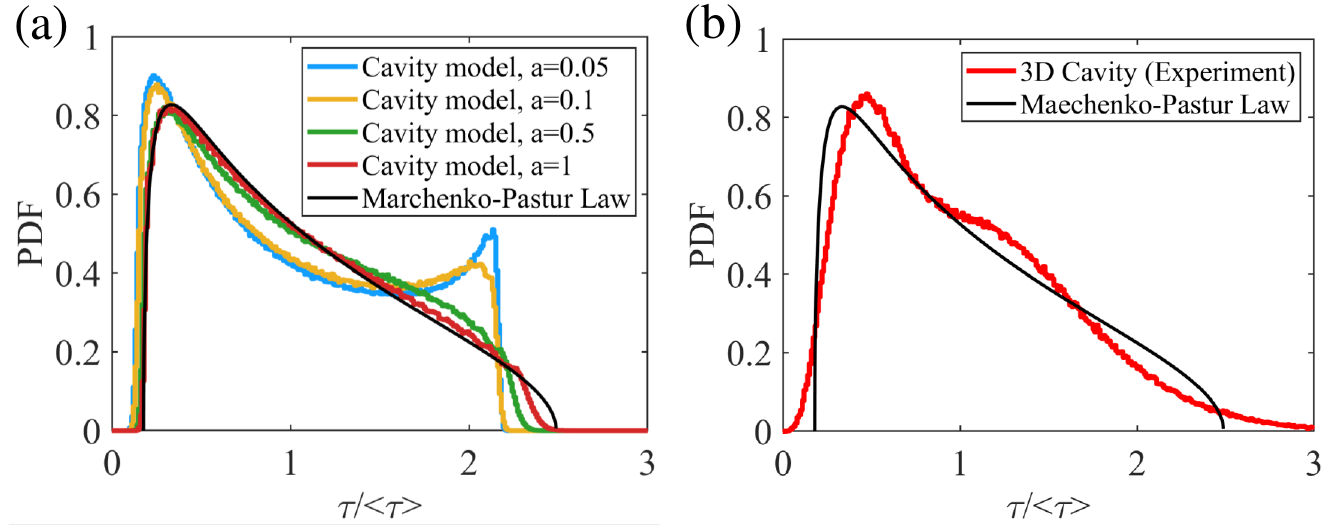}
\caption{ \label{FigS1} The PDF of $\tau/\langle \tau \rangle$ using (a) random matrix theory modeling, for cavity wave simulations, for coupling parameter $\Gamma=0.6$ and different absorption factors $a$, and (b) 3D cavity experiment (red line). The corresponding Marcenko-Pastur law is indicated in subfigures (a,b) with a black solid line. In all cases, the input/target ratio is $m_{\rm in}=1/4$, $m_{\rm tar}=3/4$.}
\end{figure*}

When the mesoscopic correlations encoded in the matrix elements of the $M_{\rm tar}\times M_{\rm in}$ rectangular matrix ${\tilde S}$ 
are washed out by excess absorption, its elements can be treated as uncorrelated identically distributed random variables with zero 
mean. In this case the probability distribution ${\cal P}(x=\tau/\langle \tau\rangle)$ of the rescaled eigenvalues of the TMT matrix 
$T$ converges to the so-called Marcenko-Pastur (MP) law when $M_{\rm in}, M_{\rm tar}\rightarrow \infty$ while keeping $M_{\rm in}/
M_{\rm tar}$ fixed. The MP law for non-zero eigenvalues takes the form
\begin{align}
{\cal P}(x) = \frac{\sqrt{(x^+ -x)(x-x^-)}}{2\pi(m_{\rm in}/m_{\rm tar})x}, 
\end{align}
where $x^{\pm}=(1\pm\sqrt{m_{\rm in}/m_{\rm tar}})^2$. This result holds for an arbitrary distribution with zero mean and equal variance of the elements of the rectangular matrix ${\tilde S}$. Hence, it also holds for the Gaussian random character expected for multiple scattered waves. The MP law has also been reproduced from our diagrammatic approach presented in Section II in the limiting case of lossy cavities ($a>1$) or strong incomplete channel control ($m_\text{in}, m_\text{tar} \ll 1$). 

Figure \ref{FigS1}(a) shows the evolution of ${\cal P}(x)$ towards the MP distribution (solid black line) in the case of chaotic cavities simulated using a random matrix theory modeling (with coupling parameter $\Gamma=0.6$), as losses are increased (the color lines indicate different $a$-values). Figure \Ref{FigS1}(b) reports the experimental results (red line) of ${\cal P}(x)$ for the reverberation cavity of Fig.~1(c) together with the MP law (solid black line). The cases depicted in Figs. \ref{FigS1}(a-b) correspond to $m_{\rm in}=1/4$ and $m_{\rm tar}=3/4$.

\begin{figure*}[t]
\centering
\includegraphics[width=\linewidth]{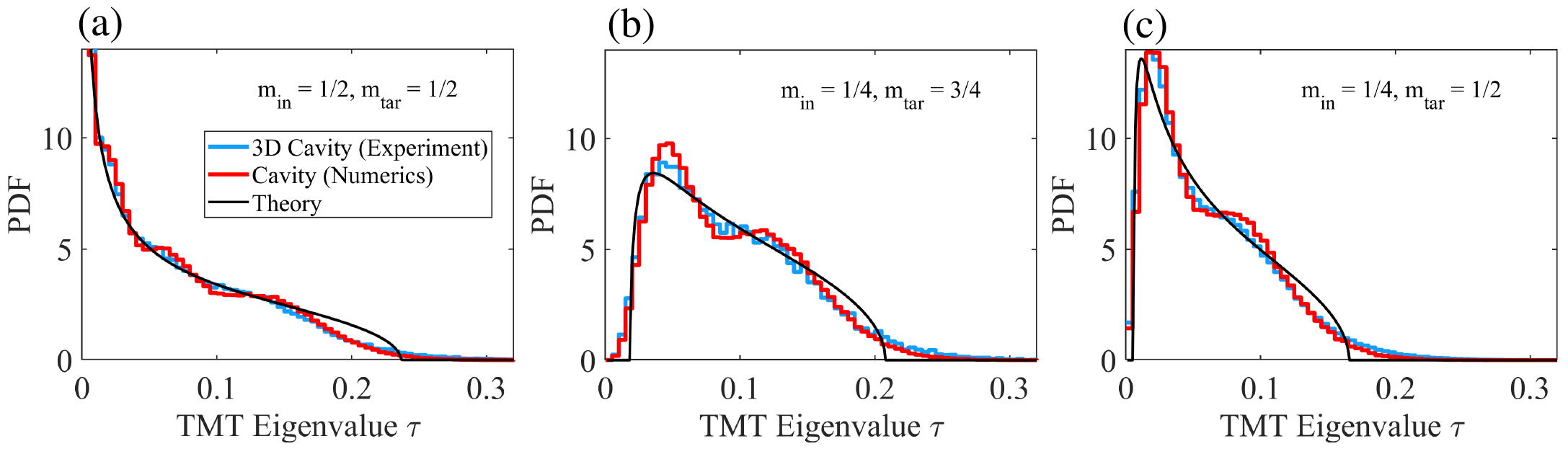}
\caption{ \label{FigS2}
The probability density function (PDF) of the TMT eigenvalues of a 3D lossy chaotic cavity. The blue, red, and black lines in each subfigure indicate the experimental measurement of 3D cavity, the cavity simulations using a random matrix theory modeling, and the predictions of the diagrammatic approach, respectively. Different input and targeted combinations are shown: (a) $m_{\rm in}=1/2$ and $m_{\rm tar}=1/2$; (b) $m_{\rm in}=1/4$ and $m_{\rm tar}=3/4$; (c) $m_{\rm in}=1/4$ and $m_{\rm tar}=1/2$. In the simulation and theory we used a coupling parameter $\Gamma \approx 0.95$ and an absorption factor $a \approx 6.3$, corresponding to a mean total absorption $A=a/(1+a/\Gamma) \approx 0.8$. 
}
\end{figure*}

Other representative $m_{\rm in}, m_{\rm tar}$ configurations are depicted in the subfigures Fig.~\ref{FigS2}(a-c). In these figures, we report the probability density function of TMT eigenvalues extracted from the measured $T$ matrices for the 3D cavity (blue lines) and the cavity simulations using a random matrix theory modeling (red lines). The corresponding diagrammatic results (which match the MP law) are also shown in these figures with black lines. In all cases, a nice agreement is observed.

\bibliography{TTM_Statistics_AM}